\documentclass[prl,twocolumn, aps,amssymb,footinbib,showpacs]{revtex4-1}

\usepackage{amsmath, amssymb, braket, dsfont, times, tensor}
\usepackage[mathscr]{euscript}
\usepackage{graphicx}
\usepackage{subfigure}		
\usepackage{epstopdf, color}
\usepackage[breaklinks]{hyperref}

\newcommand{\od}{\omega_{\textrm{D}}}
\newcommand{\oeff}{\omega_{\textrm{eff}}}
\newcommand{\td}{t_\textrm{D}}
\newcommand{\fb}{F_{\textrm{B}}}
\newcommand{\tth}{\tilde{\theta}}
\newcommand{\tJ}{\tilde{J}}

\begin{document}
\title{Classical Discrete Time Crystals}
\author{Norman Y. Yao$^{1,2}$, Chetan Nayak$^{3}$, Leon Balents$^{4}$, Michael P. Zaletel$^{5}$}
\affiliation{$^{1}$Department of Physics, University of California Berkeley, Berkeley, CA 94720, U.S.A.}
\affiliation{$^{2}$Materials Science Division, Lawrence Berkeley National Laboratory, Berkeley CA 94720, U.S.A.}
\affiliation{$^{3}$Station Q, Microsoft Research, Santa Barbara, California 93106, USA}
\affiliation{$^{4}$Kavli Institute of Theoretical Physics, University of California, Santa Barbara, CA 93106, USA} \affiliation{$^{5}$Department of Physics, Princeton University, Princeton, NJ 08540, U.S.A.}

\date{\today}							

\begin{abstract}
The spontaneous breaking of time-translation symmetry in periodically driven quantum systems leads to a new phase of matter: discrete time crystals (DTC). This phase exhibits collective subharmonic oscillations that depend upon an interplay of non-equilibrium driving, many-body interactions, and the breakdown of ergodicity.
However, subharmonic responses \cite{van1927frequency} are also a well-known feature of  classical dynamical systems ranging from predator-prey models \cite{may1976simple} to Faraday waves \cite{CrossHohenberg} and AC-driven charge density waves \cite{Brown1984}.
This raises the question of whether these classical phenomena display the same rigidity characteristic of a quantum DTC.
In this work, we explore this question in the context of periodically driven Hamiltonian dynamics coupled to a finite-temperature bath, which provides both friction and, crucially, noise. 
Focusing on one-dimensional chains, where in equilibrium any transition would be forbidden at finite temperature, we provide evidence that the combination of noise and interactions drives a sharp, first-order dynamical phase transition between a discrete time-translation invariant phase and an activated classical discrete time crystal (CDTC) in which time-translation symmetry is broken out to exponentially-long time scales.
Power-law correlations are present along a first-order line which  terminates at a critical point.
We analyze the transition by mapping it to the locked-to-sliding transition of a DC-driven charge density wave.
Our work points to a classical limit for quantum time crystals, and raises several intriguing questions concerning the non-equilibrium universality class of the CDTC critical point.

\end{abstract}

\maketitle

	Subharmonic entrainment occurs  when the long time dynamics of a system manifest a period that is a fixed multiple of the period of the underlying equations of motion \cite{van1927frequency, parlitz1997subharmonic, rasband2015chaotic, strogatz2014nonlinear}.
Such subharmonic behavior is ubiquitous in deterministic dynamical systems; most notably for example, discrete maps, e.g.~$x \to f(x)$, can exhibit stable period-doubled orbits \cite{may1976simple, linsay1981period,kaneko1984period,jackson1990periodic}.
From the point of view of many-body physics, however, in order to consider this subharmonic response  characteristic  of a phase of matter,  the system should satisfy certain properties which embody the notion of \emph{rigidity}. First,  the system should have many locally coupled degrees of freedom so that a notion of spatial dimension and thermodynamic limit can be defined. Second, the system's subharmonic response should be stable to arbitrary perturbations of both the initial state and the equations of motion, so long as the latter preserve the periodicity. Models with continuous time-translation invariance \footnote{The reader may object that the distinction between the driven and un-driven case is spurious, since we may always add an additional coordinate $\partial_t \lambda = \od$ and consider the periodic drive to be a coupling to this coordinate. However, this entails an all-to-one interaction, in contrast to the requirement of local couplings, so we feel the distinction is worth preserving.}, such as the Van der Pol oscillator \cite{van1926} and the Kuramoto model  \cite{kuramoto1975self} are not rigid in this sense, since their frequency response deforms continuously with the model parameters, although adding a small periodic drive can easily lock the response \cite{van1927frequency}.
Finally, the subharmonic response should have an infinite autocorrelation time, by analogy to ``long-range'' order \footnote{To define period doubling in certain systems (in particular, for closed Hamiltonian dynamics) some notion of coarse graining may be required. Indeed, any specific solution may be aperiodic, but an appropriate ensemble average over the initial state, spatial position, and/or time will reveal that the dynamics are period doubled in a \emph{statistical} sense.}.

\begin{figure}
\includegraphics[width=\columnwidth]{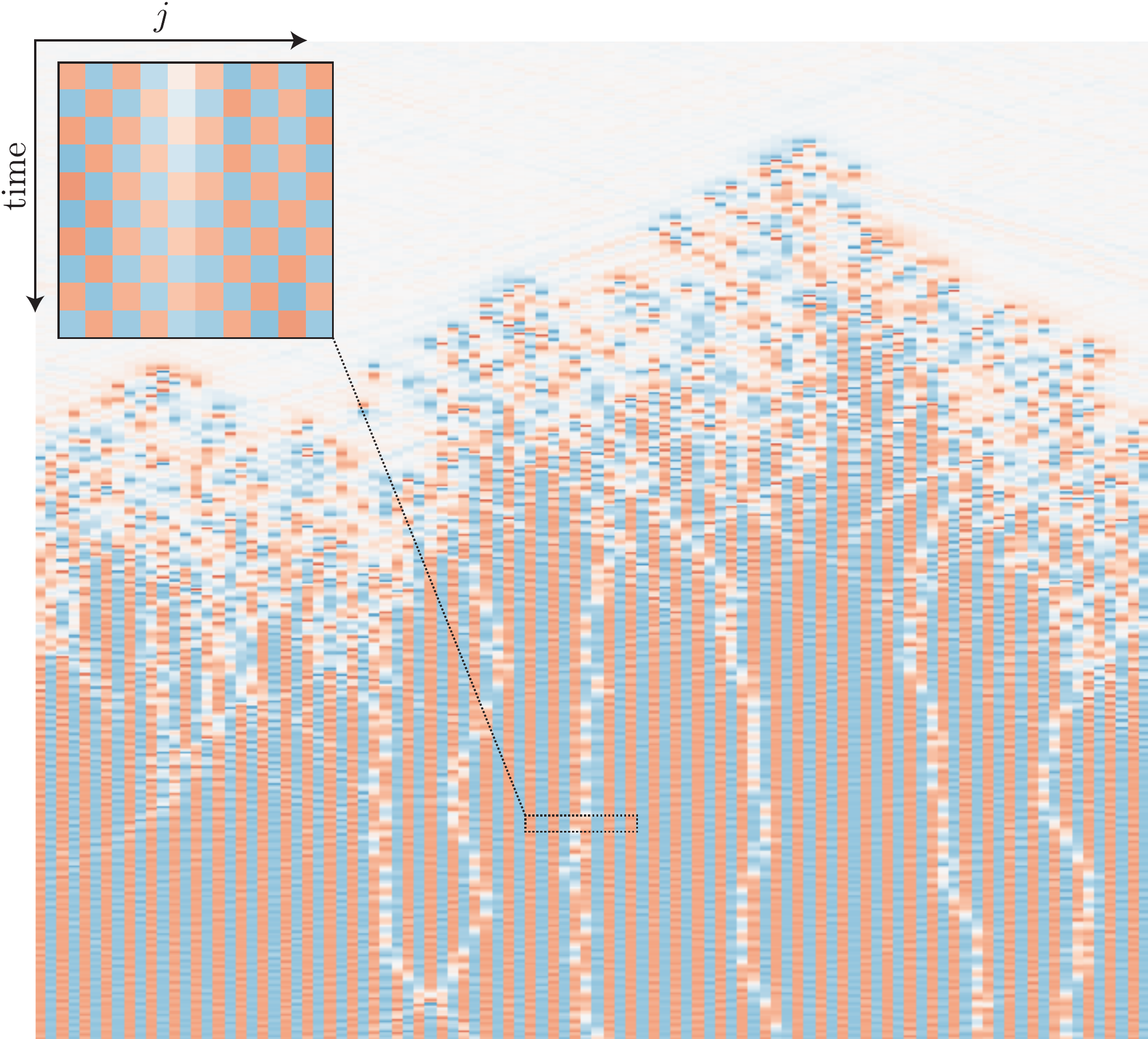} 
\caption{Period-doubled dynamics ``boil'' out of a $q_j(0) = 0$ initial state. In the main figure we present a stroboscopic view $q_j(2 n \td)$, with time $n$ running down vertically ($0 < n < 1200$) and space $j$ running horizontally over the $n_\textrm{osc} = 100$ oscillators.
The color scale is $q_j < 0$ red, $q_j > 0$ blue, and $q_j \sim 0$ white. 
Note that we strobe every \emph{two} driving periods, which is the frequency of the subharmonic response; hence the displayed phase of the oscillators varies slowly.
In the inset, we show a detail of a smaller region strobed at the driving frequency, $q_j(n \td)$.
The period-doubled oscillations $q_j(n \td) \propto (-1)^n$ are now manifest.
Strikingly, the correlations are anti-ferromagnetic both in time \emph{and} space, even though the oscillators are coupled together ferromagnetically ($\od = 1.958, g = 0.065, \delta = 0.067, \eta = 0.003, T = 0.004$).
In the final state, there is a finite density of  $\pi$-domain walls between the two different period-doubled solutions.
\label{fig:bubble}}
\end{figure}


Even within the constraints of these criteria, it turns out that general dynamical systems can still exhibit rigid subharmonic entrainment.
The reason for this is that the dynamics about fixed points can be strongly damped so that perturbations to either the state or the dynamics decay rapidly; owing to the presence of such contractive dynamics, many-body subharmonic entrainment has been observed in a multitude of systems including: Faraday wave instabilities \cite{CrossHohenberg},  driven charge density wave materials  \cite{Brown1984, brown1985harmonic, tua1985dynamics, sherwin1985complete, Wiesenfeld1987, Bhattacharya1987, Falo93, balents1995temporal} and Josephson junction arrays \cite{Lee1991, Yu1992}.

The possibility of rigid subharmonic entrainment in the absence of contractive dynamics is significantly more subtle, but also particularly relevant \cite{wilczek2012quantum,shapere2012classical,bruno2013impossibility,nozieres2013time,volovik2013broken,sacha2015modeling,watanabe2015absence}. 
Indeed, two broad classes of systems that fall into this category are time-periodic Hamiltonian dynamics in classical systems and unitary dynamics in quantum systems.
Such systems are far more restrictive than general dynamical maps; Hamiltonian dynamics, for example, are volume preserving in phase space, thereby explicitly forbidding contractive dynamics.
In the presence of an external drive (e.g.~which sets the periodicity of the dynamics), energy conservation is broken and one generically expects the long-time dynamics of the many-body system to be completely ergodic \footnote{In certain special cases \cite{citro2015dynamical,Chandran2016}, it may be possible for a driven, many-body system to avoid its ergodic fate \cite{liggett2012interacting}.}.
An ergodic system can never exhibit true subharmonic rigidity since it is impossible for the system to remember which of the distinct subharmonic orbits (i.e.~related by time translation symmetry) it began in. 

To this end, a tremendous amount of recent excitement has focused on the discovery that rigid subharmonic entrainment can occur in a periodically driven (Floquet), unitary, many-body quantum system. Dubbed Floquet/discrete time crystals  \cite{khemani2016phase,else2016floquet,von2016absolute,yao2017discrete, Khemani2017, LazaridesMoessner2017}, this new phase of quantum matter relies crucially on  many-body localization to prevent the drive-induced heating of the system to infinite temperature. 
While it is difficult to experimentally verify the long-time rigidity associated with a discrete time crystal, promising signatures of such behavior have been observed in spin systems for time-scales up to hundreds of Floquet cycles \cite{zhang2017observation,choi2017observation}.

%

A natural question thus arises: How quantum must a time crystal truly be? Is quantum mechanics important only insofar as it allows for many-body localization to prevent heating of the system?  Or does it play a more fundamental role?
If closed Hamiltonian dynamics cannot generically stabilize a time crystal due to heating,  a natural generalization is to consider preventing such heating by coupling the system to a bath,  most simply by adding friction.
However, by adding only friction, one essentially reverts back to the damped case where the existence of rigid subharmonic entrainment is well known.
On the other hand, if the bath is in equilibrium at finite temperature $T$, the fluctuation-dissipation theorem implies that  friction must come with noise \cite{nyquist1928thermal}.
In  classical systems, this noise can be captured as a Langevin force, $\fb(t) = - \eta \dot{q} + \xi(t)$, on each coordinate $q$,  where $\eta$ is the strength of the friction and $\xi(t)$ is a stochastic force with variance,  $\langle \xi(t) \xi(t') \rangle = 2 \eta T \delta(t - t')$.
Taking $\eta > 0, T=0$  reduces to the damped case where period-doubling is easily stabilized, while a combination of finite $T$ and driving results in a truly non-equilibrium situation.
The question concerning the existence of ``classical discrete time crystals'' (CDTC) can then be posed as follows: In what dimensions can a classical many-body system, coupled to an \emph{equilibrium} bath, exhibit rigid subharmonic entrainment for either the closed case ($\eta = 0$), the zero-temperature case ($\eta > 0, T = 0$), or the finite temperature case ($\eta, T > 0$)? 

One important remark. 
More generally, one can also consider coupling to a \emph{non-equilibrium} bath, which presumably leads to a larger space of stochastic systems including probabilistic cellular automata \cite{wolfram1983statistical}. In this more general context, remarkable results from G\'acs \cite{gacs1986reliable,gacs2001reliable,gray2001reader} and Toom \cite{toom1974nonergodic,toom1976unstable,toom1980multicomponent} suggest that rigid subharmonic entrainment can arise in all dimensions $d>0$ \cite{bennett1990stability}. 

%
%


\begin{figure}[t]
\hspace{-5mm}\includegraphics[width=1\linewidth]{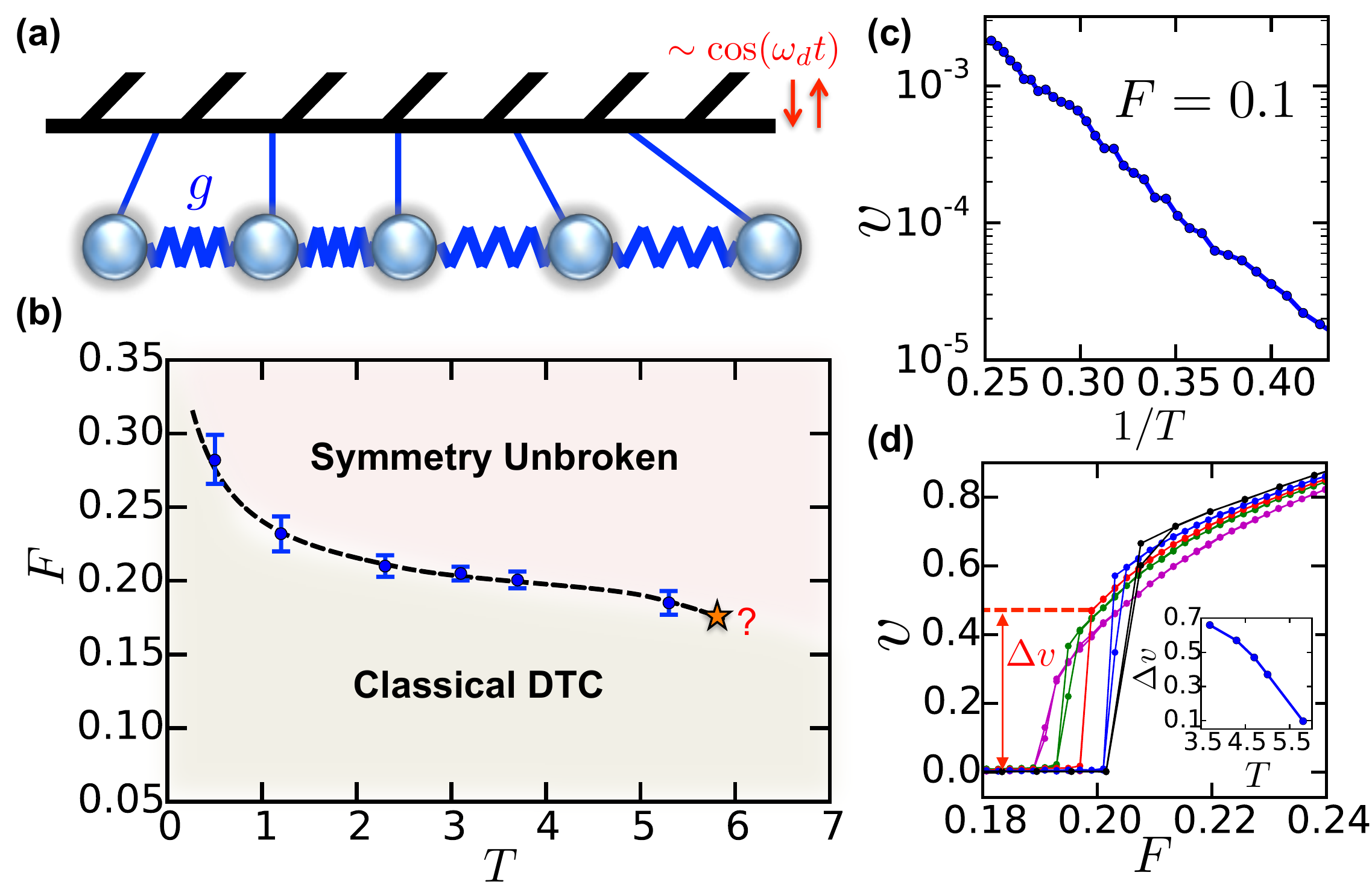}
\caption{ \textbf{a}) Schematic of a one dimensional array of coupled non-linear pendula. The pendula are coupled via ferromagnetic interactions of strength $g$, and the system is parametrically driven at frequency $\od$. \textbf{b}) Phase diagram of the classical discrete time crystal as a function of $F = \frac{2 \eta}{\delta}$ (where $\delta$ is the driving amplitude and $\eta$ the damping) and the temperature $T$. At low-$T$ there is a first order phase transition between the CDTC phase and the non-time-crystalline phase, while at high-$T$ there is only a crossover. 
\textbf{c}) In the 1D CDTC, there is  a finite rate of phase slips $v$ between the two symmetry-breaking solutions. $v$ fits very well to the Arhennius form $v \sim e^{-\Delta / T}$, indicating the phase slips are activated.
\textbf{d}) The phase transition is diagnosed by measuring the rate of phase slips $v$ between the two time-translation related period doubled solutions. 
As we cross the first order line out of the CDTC, $v$ jumps discontinuously.}
\label{fig:schema}
\end{figure}

	In this work, we focus on the finite temperature case in one dimension (1D), which is marginal in a particularly interesting sense:  while an equilibrium phase transition is impossible in 1D, might there nevertheless be a non-equilibrium dynamical phase  transition out of a period-doubled CDTC?
We find an affirmative answer, with our main results summarized as follows: first, we observe a line of first order dynamical phase transitions  between an ``activated'' period-doubled CDTC and a symmetry unbroken phase  that terminates at a critical point (Fig.~2). 
Second, we show that the phase slips associated with the transition are characterized by a $\mathbb{Z}$ index rather than a$\mathbb{Z}_2$ index. 
This implies that the dynamical phase transition we observe here is unrelated to the putative Ising transition argued to exist for isolated quantum time crystals \cite{yao2017discrete}. 
Before diving into the details, let us emphasize the subtleties which underly our results.

To begin, let us define the key characteristic of a true long-range-ordered CDTC; namely, the existence of period-doubling with an infinitely long auto-correlation time, $\tau$, which is stable to small perturbations of the dynamics.
More precisely, if one considers a  model of periodically driven oscillators with position coordinates $q_i$, the auto-correlation time, $\tau$, of period doubling can be quantified as $\langle  q_i(n \td) q_i(0) \rangle \propto (-1)^{n} e^{- n \td / \tau }$, where $\td$ is the period of the drive.
For low but finite temperatures, we observe a CDTC that exhibits an \emph{activated} auto-correlation time, $\tau(T) \sim e^{\Delta_{\textrm{eff}} / T}$, where $\Delta_{\textrm{eff}}$ is an effective activation barrier. Thus, in 1D the CDTC's time-crystalline order survives to exponentially long, but not infinite times (as we will later discuss, this caveat may be modified in higher dimensions).
In the context of experiments, it would be  challenging to distinguish this scenario from the existence of a long-range-ordered CDTC.
Despite the activated behavior of the auto-correlation time in our observed CDTC, we nevertheless find a first order dynamical phase transition where $\tau(T)$ drops discontinuously and period doubling is completely destroyed (Fig.~2). 
We hasten to emphasize that such a first order phase transition between an (activated) 1D period-doubled CDTC and a symmetry unbroken phase would be impossible in equilibrium \cite{LandauStatmech}.

\vspace{2mm}

\emph{Period doubling in the Frenkel-Kontorova model}---For concreteness, let us consider a parametrically driven Frenkel-Kontorova model, which describes an array of coupled, non-linear pendula (Fig.~2a) \cite{FK38,braun2013frenkel}: 
\begin{equation}
H =  \sum_i  \frac{1}{2} p_i^2 + \left[ 1 + \delta \cos(\od t) \right]  (1 - \cos(q_i) ) + g \sum_{\langle i, j \rangle} \frac{(q_i - q_j)^2}{2},
\end{equation}
where $q_i$ is the pendulum's deflection from vertical and $p_i$ its momentum, while $\delta$ and $\od$ are the amplitude and frequency of the parametric driving (Fig.~\ref{fig:schema}a) \footnote{Such a model can be realized in many different experimental systems. One simple circuit QED architecture is as follows. Consider a linear array of superconducting islands, with each island  coupled to a ground line via a Josephson junction with a shunt capacitance. To modulate the Josephson coupling, one can utilize a flux-tuned split junction with an oscillatory flux  generated from a nearby AC current loop. Finally, the interaction term  can be obtained by coupling each island to its neighbor with a ``superinductor''.}. Note that we have normalized each pendulum's natural frequency to one. 
When $\od \sim 2$ and $g=0$ (decoupled), each pendulum is susceptible to a 2:1 parametric resonance, where the dynamics are period doubled and the pendulum's position returns only once every two driving cycles.
While this sub-harmonic response (at $\od/2$) is reminiscent of the behavior expected for a CDTC, the parametric resonance of a single oscillator does \emph{not} exhibit the rigidity of a true time crystal; rather, as we shall see, this sub-harmonic response is destroyed as a smooth crossover for any amount of noise.
Crucially, in the presence of interactions ($|g| > 0$), the sub-harmonic response of the \emph{collective} system can undergo a sharp transition characteristic of a many-body phase.

We introduce friction $\eta$ and finite temperature $T$ through a Langevin force $F_{\textrm{B}}(t)$ which acts independently on each $q_i$.
The pendula then evolve  under a combination of this stochastic Langevin force and the ferromagnetic ($g>0$) Frenkel-Kontorova Hamiltonian:
\begin{align}
\frac{dq_i}{dt} &=  p_i \\
\frac{dp_i}{dt} &=  - (1 + \delta \cos(\od t)) \sin(q_i) \nonumber  \\
& \, \,\,\,\,\,\,\, +  g  (q_{i+1} + q_{i-1} - 2 q_i ) - \eta p_i + \xi_i(t).
\label{eq:eqom}
\end{align}
To probe the resulting dynamics, at time $t = 0$ we initialize the oscillators in $p_i(0) = q_i(0) = 0$, and integrate the equations of motion using a second-order Langevin time-stepper \cite{suppinfo}.
The  stroboscopic dynamics, $q_i(m \td)$ (where $\td = 2 \pi / \od$), are depicted in Fig.~\ref{fig:bubble}.
Strikingly, the uniform initial condition gives way to a growing bubble of spatio-temporal ``antiferromagnet'' in which $q_i(m \td) \propto (-1)^{i+m}$; these spatial anti-ferromagnetic correlations are particularly surprising, since the oscillators are ferromagnetically coupled.
The existence of a growing bubble would seem to suggest the presence of two  distinct dynamical regimes --- time crystalline and not --- despite the finite temperature fluctuations. 
 
\begin{figure}
\includegraphics[width=\columnwidth]{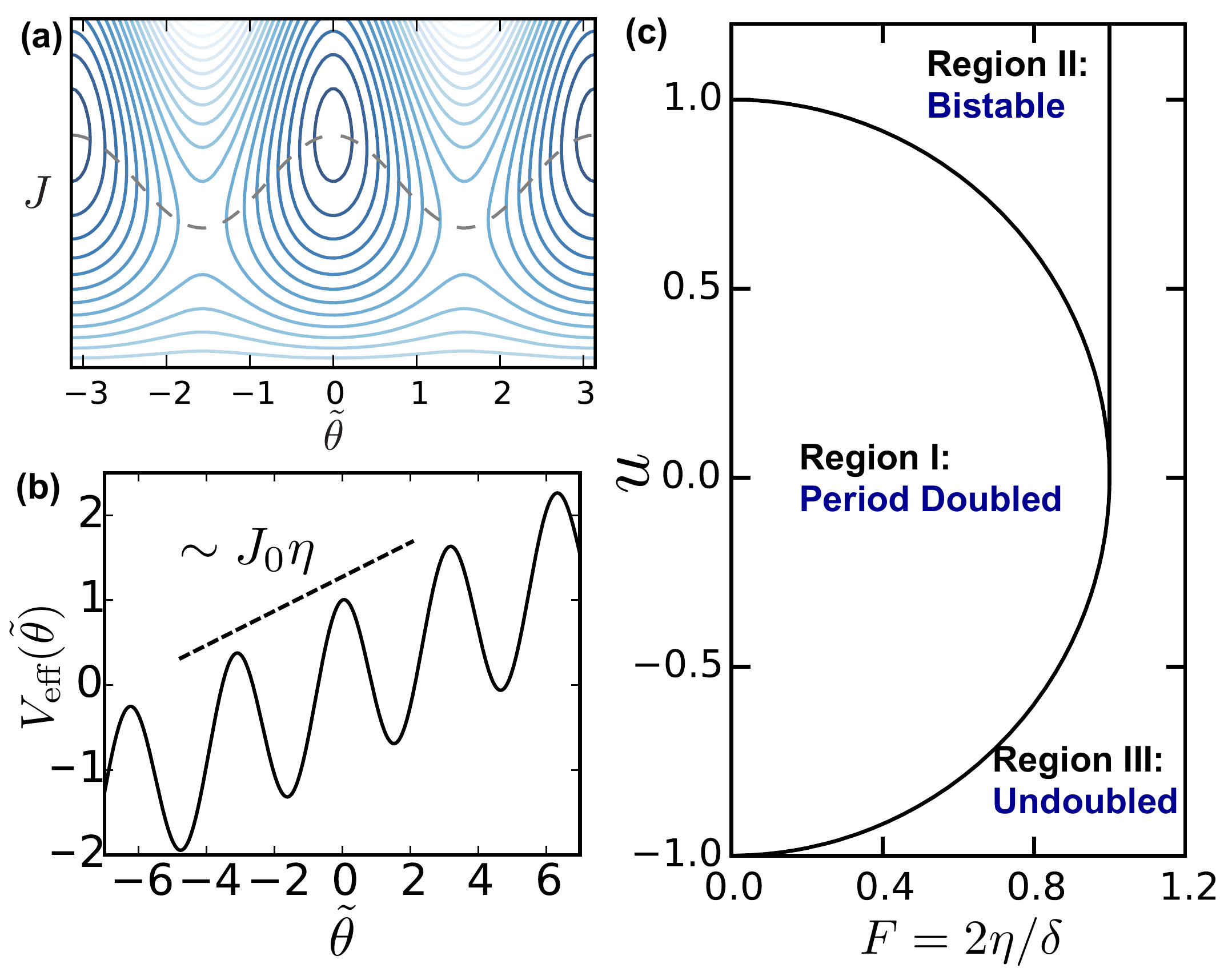} 
\caption{ \textbf{a})  Equal pseudoenergy contours of the averaged Hamiltonian $\bar{H}$ in the $\tilde{\theta}, J$ plane. The dashed line indicates the contour $\partial_{\tilde{\theta}} \bar{H} = 0$.
\textbf{b})  Effective washboard potential of Eqn.~\eqref{eq:dceff}. The slope of the potential arises from the Langevin damping $\eta$. \textbf{c}) In region I ($F > 1$ or $u < \sqrt{1 - F^2}$), the particle slides and there is no period doubling.  In Region III, $F < 1, u > 1 - F^2$, only the locked (period doubled) phase is stable. In region II, $F < 1, u^2 < 1 - F^2$, both the locked and sliding states are stable, implying bistability.\label{fig:0D}}
\end{figure}

\emph{Analysis of a single non-linear pendulum}---To begin, let us recall the parametric resonance of a single non-linear pendulum \cite{Zounes2002}. In the action-angle coordinates of the pendulum, $q \sim \sqrt{2 J} \cos(\theta), p \sim  \sqrt{2 J} \sin(\theta)$, the Hamiltonian [Eqn.~(1)] reduces to:
\begin{align}
H(t) = & J - \frac{\epsilon}{2} J^2 + \delta J \cos(\od t) \cos^2(\theta) + \cdots
\label{eq:nomassage}
\end{align}
where $\epsilon = \frac{1}{8}$ and higher-order terms in $J, \delta$ are neglected \cite{Brizard2013}. 
In the un-driven case, $\omega(J) = \partial_J H$ sets the frequency of oscillations, so that the non-linearity is encoded in  the $J^2$ term. 
Because $\epsilon  > 0$, larger  amplitude oscillations have lower frequency; however, we can be more general by keeping $\epsilon$ as a parameter and indeed will later explore the $\epsilon  < 0$ regime.

Near a period-doubled solution, $q_i \propto \cos( \od t/2 + \tth)$, where $\tth$ varies slowly. To this end, we transform to the rotating frame, $\tth = \theta - \od t / 2$, wherein Eqn.~\eqref{eq:nomassage} becomes
\begin{align}
H(t) &=  (1 - \od/2) J - \frac{\epsilon}{2} J^2 + \frac{\delta}{4} J \cos(2 \tth ) \nonumber \\
 &+  \frac{\delta}{4} J \left[ \cos(2 \od t + 2 \tth ) + 2 \cos(\od t) \right]   .
\end{align}
At leading order in a Floquet-Magnus expansion, the average Hamiltonian over a driving period is given by:
\begin{align}
\bar{H} &= \frac{1}{\td} \int_0^{\td} H(t) dt \\
&= \delta (1 + u) J/4 - \frac{\epsilon}{2} J^2 + \frac{\delta}{4} J \left( \cos(2 \tth) - 1 \right),
\label{eq:magnus}
\end{align}
where $u = 4 (1 - \od/2) / \delta$ is an effective detuning. 
The static $\bar{H}$ will govern the slow dynamics $\tth$. While this treatment is approximate (we neglect the off-resonant oscillatory terms), for a single pendulum one can in principle apply a convergent sequence of canonical transformations to bring the Hamiltonian to such a static form, a consequence of the KAM-theorem for  small $\delta$ \cite{Zounes2002}.

The equal-energy contours of $\bar{H}$ are illustrated in Fig.~\ref{fig:0D}a. For $\epsilon > 0$, the landscape is that of an inverted double well potential with \emph{maxima} at $\bar{J} = \epsilon^{-1} (1+u) \frac{\delta}{4}$ and $\tth = 0$, $\pi$, which are the two possible phase shifts of the period-doubled  solutions: $q(m \td) =  \pm \sqrt{2 \bar{J}} (-1)^{m}$. 
Since these solutions occurs at maxima, perturbations about the orbit remain bounded, oscillating at an effective frequency, $\omega_{\textrm{eff}}^2 = \delta^2 (1 + u) / 4$. 
For weak driving $\delta$, $\omega_{\textrm{eff}}/ \od$ is parametrically small, implying that higher order terms in our Floquet-Magnus expansion are strongly off resonant, justifying our approach.
We note that period-doubled solutions exist only when $\bar{J} > 0$, yielding the period-doubling criteria $u > -1$. 
Finally, for $\epsilon < 0$, the above analysis remains essentially identical except that the period-doubled solutions occur at the  \emph{minima} of $\bar{H}$.


	The  double-well potential has a  phase-shift symmetry, $\tth \to \tth + \pi$, which is unrelated to the $q \to -q$ symmetry of the original Hamiltonian, and is actually a more general consequence of  period doubling: in the rotating frame, $\tth \to \tth + \pi$ \emph{must} remain a symmetry because it corresponds to time translation symmetry, $t \to t + \td$ \footnote{In fact, the $q \to -q$ symmetry of the original Hamiltonian is inessential to the period-doubling physics we discuss, analogous to quantum case. \cite{von2016absolute}}.
	
%
	
We now come to a crucial point in the analysis. Naively, it  might seem that period doubling is analogous to the breaking of the internal Ising symmetry generated by $\tilde{\theta} \to \tilde{\theta}  + \pi $. In this case, the $\pi$-phase slips (in both time and space) between the two period-doubled solutions would have a $\mathbb{Z}_2$ character.
However, we now show that in the presence of the bath the $\pi$-phase slips in fact  have a $\mathbb{Z}$ character, implying a natural ``handedness'' to the domain walls between the two period doubled solutions.
For this reason the dynamical CDTC phase transition we observe here is not related to an Ising transition, and hence is distinct from the transition discussed in the context of the quantum MBL / prethermal time crystal \cite{yao2017discrete,ElseBauerNayak}. 
%

%
To demonstrate the $\mathbb{Z}$ character of the CDTC's  $\pi$-domain walls in the presence of damping, let us consider the equations of motion when averaging the Langevin force over one period \cite{suppinfo}:
\begin{align}
\dot{\tth} &= \partial_J \bar{H} \\
\dot{J} &= -  \partial_{\tth} \bar{H} - \eta J +  \sqrt{\bar{J}} \xi(t).
\end{align}
%
In the limit of large detuning relative to the driving ($u \gg 1$), these equation take a particularly simple form (see supplementary information \cite{suppinfo} for the general case):
\begin{align}
\label{eq:dceff}
\bar{H} &=  -\frac{\epsilon \tJ^2}{2} +\bar{J}  \left[ \frac{\delta}{4} \cos(2 \tth) + \eta \tilde{\theta}  \right]  + \cdots  \\
\dot{\tilde{\theta}} &= \partial_{\tilde{J}} \bar{H}  \\
\dot{\tilde{J}} &= -\partial_{\tilde{\theta}} \bar{H}  - \eta  \tilde{J}  + \sqrt{\bar{J}} \xi(t),
\end{align}
where $\tJ = J - \bar{J}$ and $\bar{J} = \epsilon^{-1} u \frac{\delta}{4} ( 1 + \mathcal{O}(u^{-1}) )$.
These equations describe a \emph{negative} mass particle with  ``position'' $\tilde{\theta}$ and ``momentum'' $\tJ$  subject to  Langevin damping and a washboard potential with finite slope  $\bar{J} \eta$, as shown in Fig.~\ref{fig:0D}(b).
We note that such equations of motion are extremely well studied in the context of RC-shunted driven Josephson-junctions \cite{stewart1968current, mccumber1968effect}.
Crucially, because of the bias $\bar{J} \eta \tilde{\theta}$, a phase slip of $+\pi$  is \emph{inequivalent} to a phase slip of  $-\pi$ (i.e.~one rolls down hill, while the other rolls up hill), thereby leading to ``handed'' phase slips that exhibit a  $\mathbb{Z}$ character. In the supplementary information \cite{suppinfo}, we demonstrate that this effect is ``real'' for the original driven pendulum.

	At zero damping, the barrier height of the washboard potential is $\bar{J} \delta / 2$, but as the damping $\bar{J} \eta$ increases the barrier height decreases. Thus, to parameterize the damping we consider a dimensionless ``force'', $F = \frac{2 \eta}{\delta}$, defined so that when $F \geq 1$ the stable extrema vanish and the particle slides along the washboard.
In the fully sliding state, $\dot{\tth} = \frac{\delta}{4} u $, which in the original variables gives $\dot{\theta} = 1$, e.g.~the \emph{natural}  frequency of the un-driven oscillator \cite{suppinfo}. 
Thus, the ``sliding regime'' indicates the destruction of period doubling, while the ``locked'' regime ($\dot{\tth}=0$) is period doubled.

	In the absence of noise ($T=0$), a standard stability analysis \cite{strogatz2014nonlinear} reveals three distinct dynamical regimes [Fig.~\ref{fig:0D}(c)]:  In Region I only  the sliding regime is stable, indicating there is no period doubling.  In  Region III  only the locked regime  is stable, indicating period-doubling. Finally, Region II is a bistable regime in which \emph{both} the locked and sliding states are stable, and the long time behavior depends on the initial state. This region will show hysteresis as  $F$ is varied.
	
At any finite temperature $T > 0$, activated processes cause a \emph{single} pendulum to transition between the locked and sliding states.
This destroys the bistable Region II,  leading to a smooth (though highly non-linear) crossover between the locked and sliding regimes as either the force ($F$) or temperature ($T$) is increased \cite{buttiker83}.
This crossover in the effective static model is consistent with our numerical experiments on the original parametrically driven pendulum [Eqn.~\ref{eq:eqom}]. Thus, for $g=0$ (i.e.~in 0D) and at finite-$T$ there is no sharp transition between  period-doubled and undoubled dynamics.

\emph{Collective behavior of  coupled non-linear pendula}---With the single pendulum analysis behind us, let us now 
turn to the collective behavior of the system at finite coupling strength. 
We begin by noting that at finite damping and zero-temperature ($\eta > 0, T=0$), the 1D  chain will exhibit stable many-body period doubling which descends  from the single oscillator case \footnote{Roughly speaking, if we perturb about the period-doubled solution of a single oscillator $q(t)$,  $q_j(t) =  q(t) + \Delta q_j(t)$, in the limit $\delta \to 0$ the dispersion relation of the perturbation is strongly off-resonant with the drive, $\omega_{\textrm{eff}}(k) / \od \sim \delta$, so the finite damping $\eta$ prevents heating and the system settles into a period-doubled steady state.}.
For finite temperatures ($T>0$), one might expect the coupled  chain to exhibit only a smooth crossover as $F$ and $T$ are varied, 
 analogous to the 0D case, since one is tempted to think of a potential transition as a finite-$T$ equilibrium Ising transition,  which cannot exist in 1D.
However, the appearance of the effective bias $\bar{J} \eta$ highlights that the system is intrinsically non-equilibrium, so  such constraints need not  apply.
Indeed, there is evidence that the 1D DC-driven Frenkel-Kontorova model shows a first-order locked to sliding transition at finite temperature \cite{braun1997nonlinear, braun1997hysteresis, braun2013frenkel}.
 
%
%

As before, to analyze the interactions within an effective static model we average the couplings over one Floquet period, $\bar{H}_{g} = - g \sum_{i} \sqrt{J_i J_{i+1}} \cos(\tth_i - \tth_{i+1})$ \cite{suppinfo}, leading to the effective  equations of motion:
\begin{align}
\label{eq:dceffmultiple}
\bar{H} &=  \sum_i \biggl ( -\frac{\epsilon \tJ_i^2}{2} + \bar{J} \left[ \frac{\delta}{4} \cos(2 \tth_i) + \eta \tth_i  \right]  \\
& \quad  \quad  \quad - g \bar{J}  \cos(\tth_i - \tth_{i+1}) \biggr )  \\
\dot{\tth}_i &= \partial_{\tilde{J}_i} \bar{H}  \\
\dot{\tilde{J}}_i &= -\partial_{\tth_i} \bar{H}  - \eta  \tilde{J}_i  +  \sqrt{\bar{J}} \xi_i(t),
\end{align}
where we have taken $J_i \approx \bar{J} + \mathcal{O}(1/u)$. Except for the negative mass and the finite slope $\eta \tth$, these equations correspond to the sine-Gordon representation of the Ising model.

Before exploring the CDTC phase transition in this model, a few remarks are in order. 
First, the negative mass explains why the ferromagnetically coupled pendula displayed  \emph{anti}-ferromagnetic spatial synchronization in Fig.~\ref{fig:bubble}. 
Because the period doubled orbit of a single pendulum occurs at a maximum of $\bar{H}$, the volume of available phase space increases as the quasi-energy decreases. 
Interpreting this as a relation between entropy and energy, the period-doubled solution is at negative temperature, even though the Langevin bath is at positive temperature.
Thus, the array is entropically driven toward a \emph{high} quasi-energy state, reversing the expected effect of the coupling $g$ \footnote{To test this hypothesis, we can change the effective mass to be positive  by changing the non-linearity $\epsilon$ of the pendulum (e.g., by replacing $\cos(q) \to \frac{1}{2} q^2 + \frac{1}{24} q^4$, giving $\epsilon = -\frac{1}{8}$).
The period doubled solutions now exist at minima of $\bar{H}$, corresponding to positive temperature, and analogous simulations indeed show that the pendula now synchronize ferromagnetically.}.
Second, while negative temperatures are familiar in models with finite phase space, the parametric resonance of a non-linear pendulum provides a novel way to dynamically generate negative temperatures in a system with an unbounded phase space. 
We note that this phenomena is  quite distinct from the dynamically stabilized inverted position of a Kapitza pendulum, which remains at a minimum of the quasi-energy; indeed, ferromagnetically coupled Kapitza pendula will synchronize ferromagnetically \cite{kapitza1951} .

\emph{CDTC phase transition}---While $\epsilon > 0$ provides access to an intriguing negative temperature regime, from the perspective of time translation symmetry breaking, the sign of $\epsilon$ does not appear to impact the CDTC. Thus, for the sake of simplicity and to simplify the visual presentation, we will utilize the following potential: $\frac{1}{2} q^2 + \frac{1}{24} q^4$
(leading to $\epsilon = -\frac{1}{8}$ and hence, ferromagnetic spatial synchronization) for the remainder of the text. 

In the absence of a slope in the washboard potential, the system is equivalent to an equilibrium 1D Ising model, implying that at finite temperature, there will always be a finite density $\sim e^{ - g_{\textrm{eff}} / T}$ of $\pi$-domain walls in space, as well as a finite rate of $\pi$-phase slips $\sim e^{- \Delta_{\textrm{eff}} / T}$ in time, where $g_{\textrm{eff}}, \Delta_{\textrm{eff}}$ are effective quasi-energy barriers. 
This is the reason why in one dimension, one expects that the CDTC phase will exhibit an  ``activated'' autocorrelation time: $\tau \sim e^{\Delta_{\textrm{eff}} / T}$.
However, because of the finite slope $\eta \tth$ in the washboard potential, a more basic question is whether the period doubling is  completely destroyed by a collective sliding state.
Unlike in 0D, the coupled chain may exhibit a sharp transition between the locked and sliding states even at finite temperature \cite{braun1997nonlinear, braun1997hysteresis},  implying a non-equilibrium phase transition between an activated CDTC and a symmetry unbroken state.

 
 \begin{figure}
\hspace{-6mm}\includegraphics[width=3.6in]{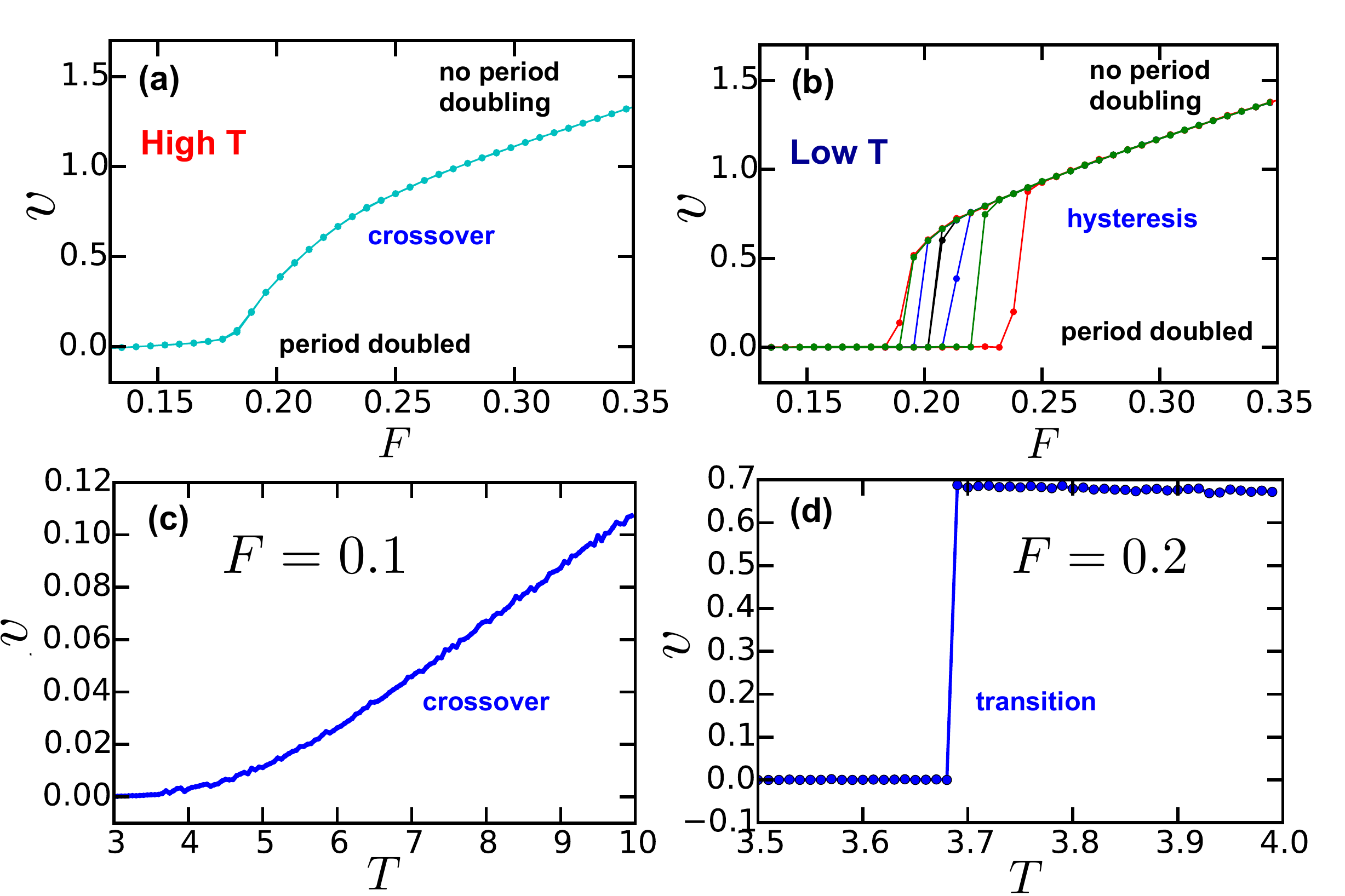}
\caption{Rate of phase slips $v = \langle \dot{\tth} \rangle$ as a function of the damping $F$ and temperature $T$. \textbf{a})  High temperature ($T=7$).
For low damping $F$, $v \sim 0$ indicating period doubling, while at large $F$, $v>0$ indicates that a finite rate of $\pi$-phase slips destroys true long-range order in time. There is a smooth crossover between these two regimes, with no evidence of hysteresis. \textbf{b}) At lower temperatures, ($T=3.7$), we observe a  hysteresis loop that closes into a jump discontinuity as we increase the simulation time. Curves of different color correspond to simulations of length $n_\textrm{step} \in$ (5k, 30k, 300k, 3m) for each value of $F$ as we sweep up and down. This suggests a first-order non-equilibrium phase transition. \textbf{c}) Analogously, we can hold $F$ fixed and vary $T$. At $F = 0.1$, $v$ displays  a crossover, while at  $F=0.2$, \textbf{d}), there is a sharp transition. Repeating this analysis throughout the $F, T$ plane allows us to trace out a first-order line which terminates, yielding the phase diagram in Fig.~\ref{fig:schema}(b).}
\label{fig:v}
\end{figure}

	To investigate this possibility numerically, we explore the system's behavior as a function of both temperature, $T$, and the dimensionless ``force'', $F$, associated with the washboard slope. 	
We fix $\eta = 0.005$ so that $\delta = 2 \eta / F$, and for each $F$ we adjust $\od, g$ to  keep fixed $\oeff = 0.1$ and $g = 1.25 \delta$.
This choice ensures that throughout the phase diagram we explore  $u \gg 1, F < 1$, so that the oscillators are individually in the bistable regime (Region II).

A quantitative diagnostic of period doubling is given by the ``velocity'' parameter, $v = \langle \dot{\tth} \rangle$, where the average  is taken  over time, space, and different realizations of the stochastic force. 
Period doubling corresponds to the locked state where $v=0$, while $v \neq 0$ (sliding state) implies a finite rate of phase slips. 

Several representative cuts of $v(F, T)$ in the $(F, T)$-plane are depicted in Figure~\ref{fig:v}, which exhibit two regimes. 
At high temperatures (Fig.~\ref{fig:v}a), $v(F, T)$ varies smoothly with $F$ and no transition is observed, similar to the zero dimensional case of a single pendulum.
However, at low temperatures (Fig.~\ref{fig:v}b), $v(F, T)$ displays hysteretic behavior as one sweeps the force.
By increasing the time scale of each sweep (e.g. increasing the number of driving periods at each $F$ from $n_{\textrm{steps}} = 5 \times  10^3$ to $3 \times 10^6$), the hysteresis loop closes into a single-valued curve that exhibits an apparent  jump discontinuity. 
This suggests that the interactions have transformed the bistable Region II of the individual pendula  into a finite-temperature first order dynamical phase transition of the coupled chain.
As an additional test, one can fix $F$ to its value near the transition and slowly vary $T$, reproducing the same  discontinuous jump (Fig.~\ref{fig:v}d).

%

\begin{figure}[t]
\includegraphics[width=1.\linewidth]{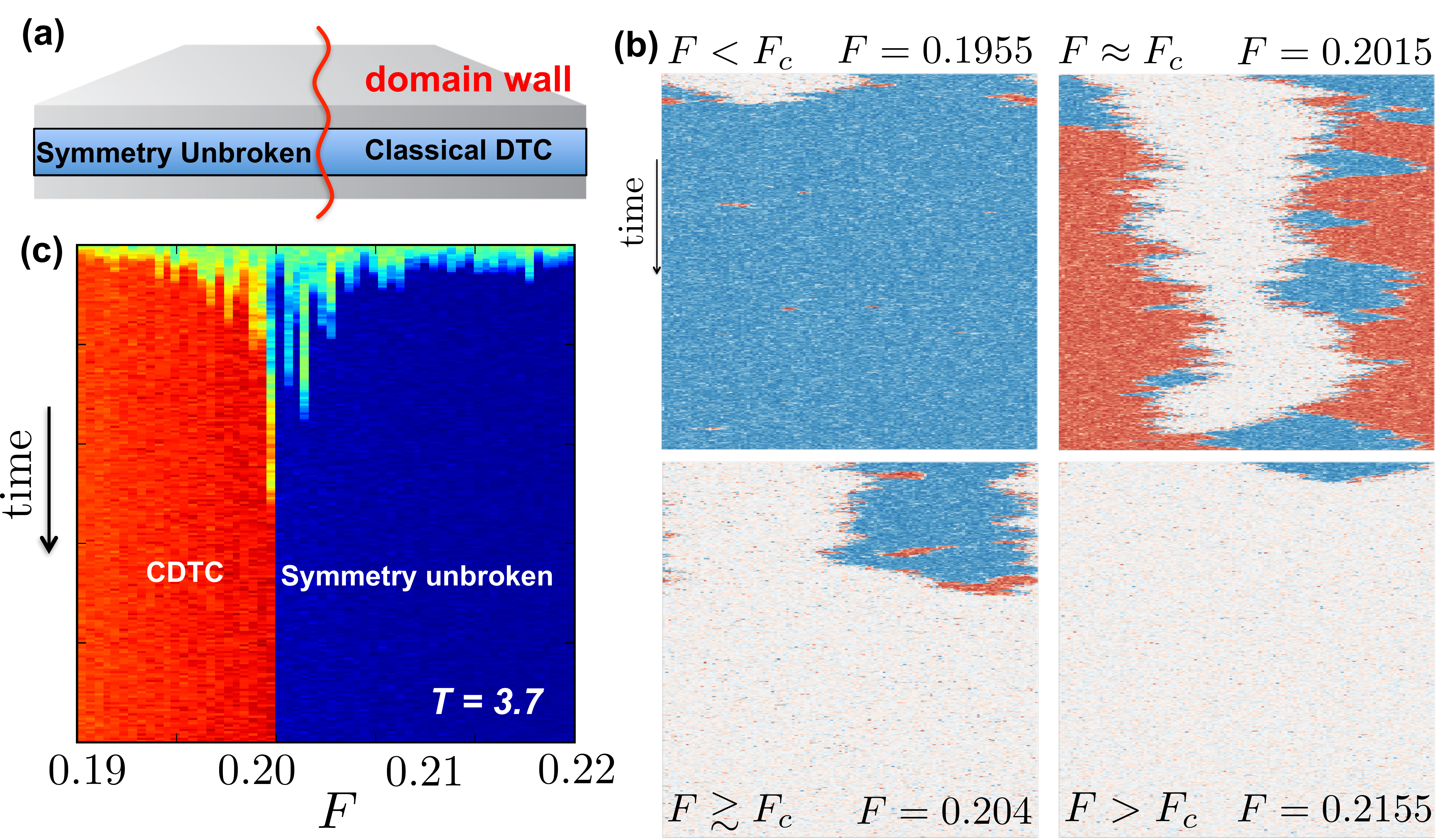}
\caption{Competition between period doubled and undoubled dynamics near the putative first order transition. By introducing a dynamical domain wall between the two regimes by hand, \textbf{a}) we circumvent the exponentially large time scale which leads to hysteresis.
Similar to Fig.~\ref{fig:bubble}, in \textbf{b}) we present $q_j(2 n \td)$ for $n_\textrm{osc} = 10^3$ oscillators.
We strobe every \emph{two} driving periods to avoid plotting the period-doubled oscillations. 
Thus the red and blue regions  indicate one or the other period doubled orbits, while in white regions there is no period doubling.
At $t=0$, we initialize a DDW using an initial state in which either $q_j(0) = 0$ (left half) or $q_j(0) \sim \sqrt{2 \bar{J}}$ (right half), and then evolve for $n_{\textrm{steps}} = 10^6$ periods. If a first order transition exists at critical force $F_\textrm{c}$, we expect a sensitive dependence of the DDW dynamics on $F$ near the transition.
Indeed, for $F < F_\textrm{c}$,  the period doubled region expands and ``eats'' the non-time-crystalline region, while for $F > F_\textrm{c}$, we see the opposite. Meanwhile,  close to the transition point, $F_\textrm{c} \sim 0.2015$, the competition between the two phases extends for many steps. To locate the transition precisely, in \textbf{c}) we use a color plot to display the time evolution of the average oscillation amplitude $\langle J(t) \rangle_{F}$ during the DDW quench. Since $\langle J \rangle$ differs in the two regimes, the long time behavior converges to one of two possible values as one or the other domain ``wins''. 
}
\label{fig:domainwars}
\end{figure}


	While the jump looks sharp to the eye, it is difficult to numerically locate the transition in this manner, because the time required to close the hysteresis loop diverges at low temperature.
To ameliorate this issue, we study the behavior of a ``dynamical'' domain wall (DDW) between the period-doubled and un-doubled states, since presumably it is the nucleation of the first such DDW that  requires the largest time  \cite{dhuse}.
Specifically, we initialize the left half of the system to be in the symmetry unbroken state with $q, p \sim 0$ and the right half of the system to be in the period doubled CDTC with $q, p \sim \sqrt{2 \bar{J}}$ (Fig.~5a).  We then time evolve for $n_{\textrm{steps}} = 10^6$ to determine which state ``wins.''
In Figure~\ref{fig:domainwars}(b,c), we fix $T = 3.7$ and repeat this experiment for a very narrow window of $F$ around the putative transition at $F_c$.
For $F < F_c$, we observe the CDTC region expand and ``eat'' the non-time-crystalline region, while for $F > F_c$, we see the  opposite behavior. 
Meanwhile, close to the critical point, $F_c$, the competition between the two phases extends for many steps as the location of the DDW fluctuates, indicating coexistence \cite{huse1987dynamics}.

To quantify this competition between the two domains, we measure the average oscillator amplitude, $\langle J(t) \rangle$, as the system evolves after the quench [Fig.~\ref{fig:domainwars}(c)].  
Since $J$ differs between the CDTC and symmetry unbroken states,  its spatial average indicates which domain is winning (although other local observables would serve just as well).
%
 %
 Far from $F_c$, $J$ converges rapidly in time to the value it takes in either the CDTC ($F < F_c$) or symmetry unbroken  ($F > F_c$) state, indicating that one or the other domain has taken over.
As  $F \to F_c$, the time required for convergence increases, and we utilize the long time behavior  to accurately determine $F_c \approx 0.201$. 
One expects that for larger systems, this convergence time scale will diverge, owing to the diffusive dynamics of the DDW, although we have not investigated this quantitatively \cite{huse1987dynamics}.

By repeating this analysis as a function of $F$ and $T$, we obtain the CDTC phase diagram  depicted in Fig.~\ref{fig:schema}(b).
We observe a line of first-order dynamical phase transitions terminating at a point in the $(F, T)$ plane.
As expected, in the CDTC region of the phase diagram, the rate of phase slips exhibits activated behavior: $v \sim e^{-\Delta_{\textrm{eff}}(F) / T}$ [Fig.~\ref{fig:schema}(c)], while at the first order transition, $v$ jumps discontinuously into a regime with  \emph{complete} destruction of  period doubling.
%
%

The nature of the end-point of the first order line is an intriguing question for future study. 
In Fig.~\ref{fig:schema}(d), we show the magnitude of the jump discontinuity, $\Delta v$, across the first order line for a range of temperatures. 
The magnitude of the jump decreases as we approach the end-point of the first order line, consistent with a scenario in which the phase transition becomes continuous at the critical end-point.
%
%
Understanding this critical point would be a fruitful starting point for a field-theoretic understanding of the CDTC transition.

 \begin{figure}[t]
\includegraphics[width=1.05\linewidth]{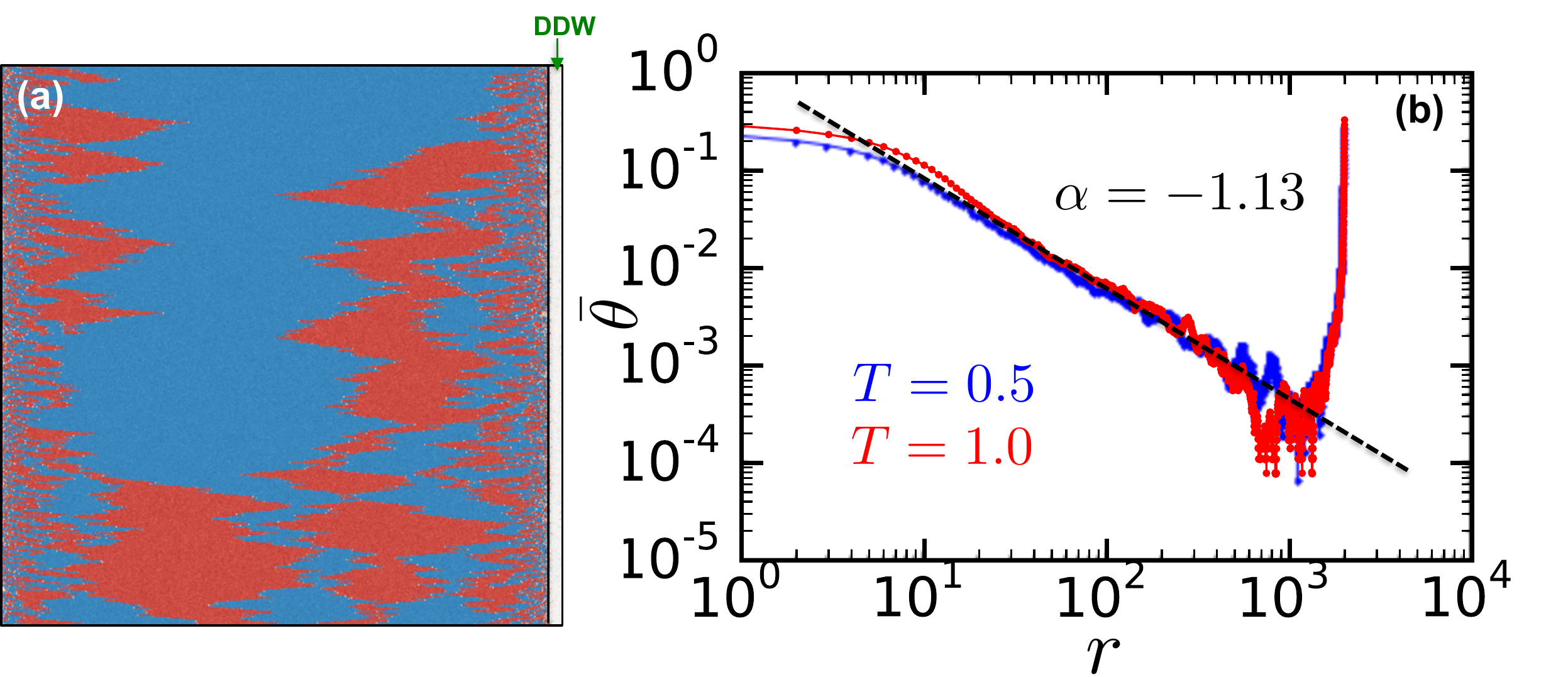}
\caption{
One of the most striking features of the first order transition is the presence of power law correlations.
 \textbf{a}) To probe these correlations, we pin a pair of dynamical domain walls by spatially modulating $F$ above and below $F_\textrm{c}$ (note the small white strip on the right, which is the high-$F$ region), and observe the resulting dynamics $q_j(2 n \td)$.
The DDW emits a constant stream of $\pi$-domain walls into the CDTC region, visible as boundaries between red and blue domains. Their density is quantified by the average velocity $ v(r) = \langle \dot{\tilde{\theta}}(r) \rangle$.  \textbf{b}) The density of $\pi$-domain walls follows a power law $v(r) \sim r^{-\alpha}$ away from the DDW, where  $\alpha \approx 1.13$. The exponent appears to be constant over several temperatures (here $T = 0.5, 1$). }
\label{fig:powerlaws}
\end{figure}

If the coarse-grained behavior of the DDWs were governed by an effective free-energy functional with short-range interactions, then   entropic arguments would imply that the putative transition is in fact rounded-out to a crossover \cite{LandauStatmech}. However, far from equilibrium, it is unclear that a free energy based argument has any relevance. Moreover, a non-equilibrium system can generically develop power-law correlations which may mediate power-law interactions between the DDWs.
To explore this, we pin a pair of DDWs by spatially modulating the force slightly above and below its critical value.
As depicted in Fig.~\ref{fig:powerlaws}(a), we find that the DDW boundary between the CDTC and symmetry-unbroken regions emits a finite density of $\pi$-phase slips. 
These $\pi$-phase slips contribute to a finite velocity, $v(r)$, that depends on the distance $r$ from the DDW. 
We observe $v(r) \sim r^{-\alpha}$ over almost two decades, where $\alpha \sim 1.13$ [Fig.~\ref{fig:powerlaws}(b)]; within the accuracy of our numerics, we obtain the same exponent $\alpha$ for cuts across the transition at two different temperatures $T = 0.5, 1$.
This power law behavior is certainly distinct from the expectations for an equilibrium first order transition.

\begin{figure}[t]
\includegraphics[width=1.0 \columnwidth]{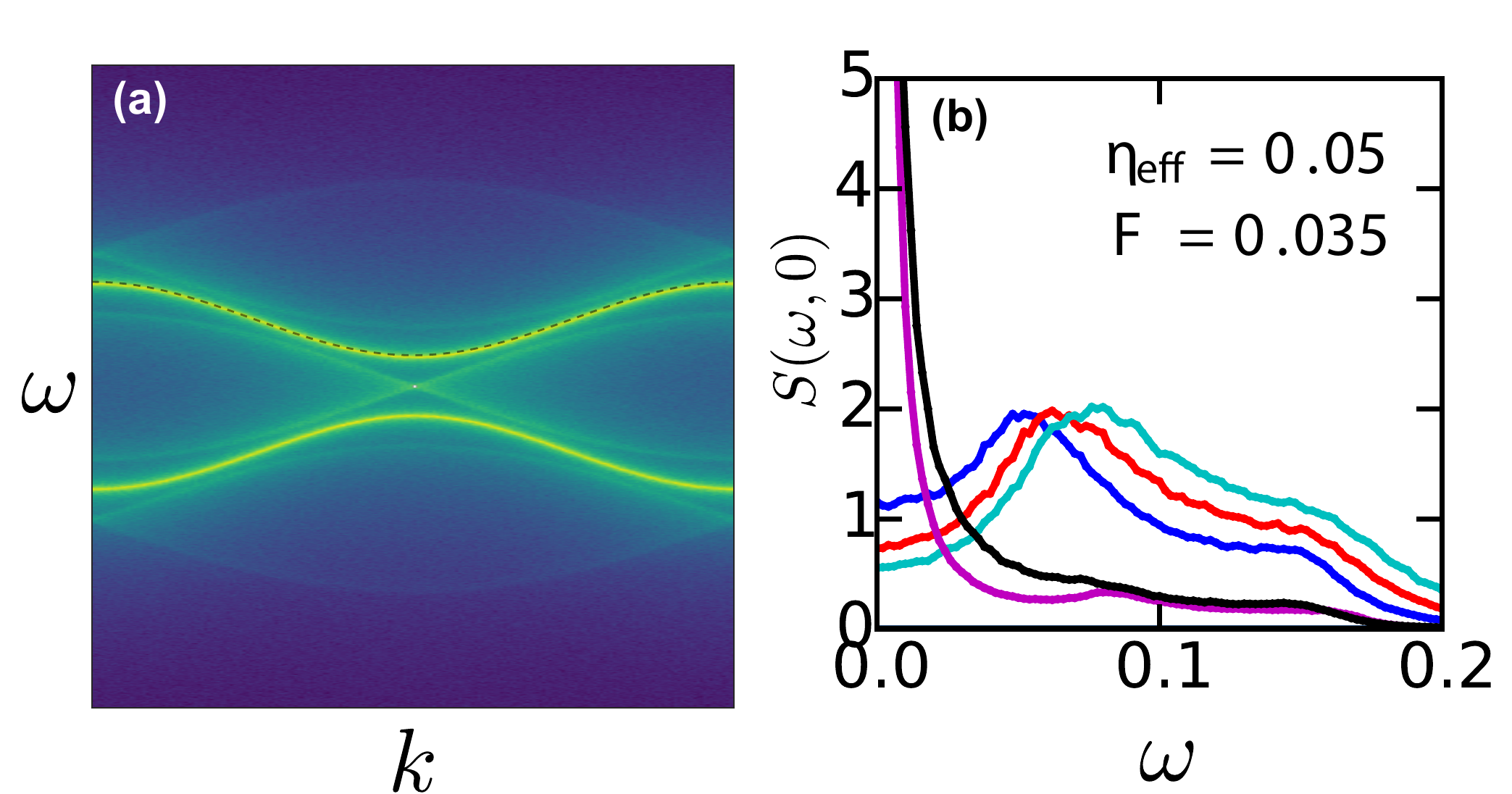}
\caption{\textbf{a})~The stroboscopic spectral function $S(\omega, k) \propto  \langle |\textrm{DFT} \left[ (-1)^n q_j(n \td ) \right] |^2 \rangle$. In this convention, the period doubled component is mapped to $\omega=k=0$. Since the autocorrelation time is unmeasurable over the $n_\textrm{steps} = 15000$ used to take the data, there is a $\delta$-function peak at the origin we have removed by hand to  preserve the scale. The residual noise spectrum reveals the mode $\omega_{\textrm{eff}}(k)$ of the effective Floquet-Magnus Hamiltonian Eqn.~\eqref{eq:dceffmultiple}. \textbf{b}) $S(\omega, 0)$ as a function of frequency at several temperatures, holding the other parameters (e.g., $F, \eta, g$) fixed. The lowest temperatures (purple, black) lie below the first-order transition, $T < T_c$ and the peaks (if resolved) would exhibit a very small  shift away from $\omega = 0$ due to the exponentially-rare activated phase slips. For the remaining curves $T > T_c$, and the peaks are strongly shifted away from $\omega = 0$ indicating unlocking of the subharmonic response. } 
\label{fig:Swq}
\end{figure}

Finally, we will now consider a diagnostic of the CDTC phase which is amenable to experiments, namely,  the power spectrum of  $q_j$. 
In particular, let us begin by defining the ``stroboscopic'' Fourier transform: $q(\omega, k) \equiv \sum_{n, j} (-1)^n e^{i (\omega n - k j) }q_j( n t_\textrm{D})$ \footnote{An analogous analysis would apply to a continuous-time Fourier transform.}. 
In order to estimate the power spectrum,  $S(\omega, k) =  \langle q(-\omega, -k) q(\omega, k) \rangle / n_{\textrm{osc}} n_{\textrm{steps}}$, we utilize Welch's method and  average over the stochastic noise \cite{welch1967use}.
 A typical spectral function is illustrated in Figure~\ref{fig:Swq}(a) and reveals the dispersion relation of the effective Hamiltonian in  Eqn.~\eqref{eq:dceffmultiple}.

 For $k=0$, perfect period-doubling would manifest as a $\delta$-function peak at $\omega = 0$.  A shift in the peak (away from $\omega = 0$) indicates  unlocking from the subharmonic response $\od/2$.
Since  $(-1)^n q( n \td )  \sim \sqrt{2 \bar{J}} \cos( \tilde{\theta}(t))$, this shift  is precisely the velocity, while the broadening is analogous to a Debye-Waller factor.
In Figure~\ref{fig:Swq}(b), we depict $S(\omega, 0)$ for a range of temperatures across the  phase transition and observe  a qualitative transformation below $T_c$.
As the temperature  decreases even further, the spectral function has to be averaged over extremely long times in order to detect the exponentially small shift and broadening of the peak away from $\omega = 0$; nevertheless, it can be used to experimentally detect the first-order jump in $v$ shown in  Fig.~\ref{fig:v}(d).

\emph{Discussion}---We have demonstrated that a periodically driven, one-dimensional system at finite temperature can exhibit a first-order dynamical phase transition between an activated  classical discrete time crystal and a symmetry-unbroken phase.
This behavior depends crucially on the interplay of interactions and non-equilibrium driving, without which a transition would be  forbidden.


	Our work opens the door to a number of intriguing future directions. First, while the 1D CDTC model studied here is thermally activated, it would be useful to rigorously establish whether a true CDTC is generically impossible in 1D. An infinite auto-correlation time requires ergodicity breaking, usually thought to be impossible in 1D due to the ``positive rates conjecture''. The positive rates conjecture roughly states that a local stochastic process in 1D is generically ergodic, generalizing the folklore regarding the impossibility of ferromagnetism is 1D \cite{liggett2012interacting}. However, a remarkable counter-example to this conjecture was provided by G\'acs in the form of a stochastic 1D cellular-automaton \cite{gacs1986reliable, gacs2001reliable,gray2001reader}. This counter-example may allow one to construct a true 1D CDTC in such models despite the noise. However, it is unclear if our restriction to continuous time-periodic Hamiltonian Langevin dynamics admits a similar construction.

	In $d>1$, where ergodicity can certainly be broken, the possibility of a true CDTC is even more complex. This issue was considered in the context of stochastic cellular-automata, where it has been argued that a higher-order subharmonic response  (e.g. periodic-tripling, $k=3$) cannot have an infinite auto-correlation time, while a period-doubled response could \cite{bennett1990stability}. The basic argument is that if a fluctuation nucleates a bubble of the neighboring subharmonic orbit (the generalization of our $\pi$-phase slips), the domain wall will generically experience a force which causes the bubble to expand. They argue that this force is absent for $k=2$ because an Ising-like domain wall does not have an orientation, leading to the distinction between $k=2$ and $k > 2$.

	Our mapping to a tilted sine-Gordon model provides a new perspective on their analysis. For a $k$-th order subharmonic response, one can generalize our effective Hamiltonian  by simply replacing the period of the washboard potential with $\cos(k \theta_j)$. As we have emphasized,  when the domain wall is smooth it still has a definite handedness even when $k=2$, so the distinction between $k = 2, k > 2$ may not seem so important. Indeed, in the continuum limit, $k$ drops out of the resulting sine-Gordon model. For a tilted sine-Gordon model satisfying detailed balance, the ``locked'' phase, with true long-range order at the subharmonic period, is unstable in any dimension: for any non-zero tilt there is only a finite energy barrier to create a phase slip domain, which, once nucleated, grows rapidly to infinite size due to the tilt. However, nucleation could in principle be prevented by non-equilibrium effects. Furthermore,  the locking to a commensurate frequency and the existence of spontaneous oscillations are potentially distinct: Even the ``sliding'' phase may exhibit long-range or quasi-long-range \cite{grinstein1993temporally} order at a frequency {\em shifted} from $\od/2$.   Such ``incommensurate'' temporal order was discussed in the context of driven periodic media \cite{balents1995temporal}. 

	Furthermore, while the handedness (and hence force) on domain walls is well defined in the sine-Gordon continuum limit, the lattice allows for $2 \pi$-vortices  at which the orientation of the domain wall reverses. These vortices cost finite energy when localized to the domain wall, where they may proliferate due to fluctuations. This may provide a mechanism which renormalizes the force on the domain walls to zero, stabilizing a true CDTC for $k=2$ and $d > 1$. 

Second, another interesting scenario is to consider the $\eta, T \to 0$ limit of our model, which reduces to closed Hamiltonian dynamics. 
Here, we observe that the dynamics remain  period-doubled out to extremely long time scales (e.g. many millions of driving cycles). 
This behavior appears to be the classical analog of a ``prethermal'' time crystal and arises from a mismatch between the driving frequency $\od$ and the \emph{effective} resonance frequency $\oeff$ \cite{citro2015dynamical,ElseBauerNayak}. 
 In the quantum case, prethermalization has a rigorous mathematical underpinning \cite{Abanin2017, ElseBauerNayak} and it would be insightful to develop its classical analog.

Finally, the existence of a first-order line  terminating at a critical point is reminiscent of the equilibrium  liquid-gas transition. 
At the critical point, one could develop a non-equilibrium field theory for the transition  within  the Martin-Siggia-Rose path integral formalism \cite{MartinSiggiaRose}. 
Such a field theory might also be used to determine the precise relation  between classical and quantum time-crystals by analyzing the semiclassical limit of its Keldysh path integral.

\begin{acknowledgments}
We gratefully acknowledge the insights of and discussions with E. Altman, D. Huse, S. Gazit, L. Sieberer, S. Sondhi, and B. Zhu. This work was supported in part, by the NSF (PHY-1654740), the ARO (W911NF-17-1-0606), the DOE, and the LDRD program of LBNL. LB was supported by the NSF Materials Theory program through grant DMR1506119.
\end{acknowledgments}

\bibliography{TC}

\begin{thebibliography}{75}%
\makeatletter
\providecommand \@ifxundefined [1]{%
 \@ifx{#1\undefined}
}%
\providecommand \@ifnum [1]{%
 \ifnum #1\expandafter \@firstoftwo
 \else \expandafter \@secondoftwo
 \fi
}%
\providecommand \@ifx [1]{%
 \ifx #1\expandafter \@firstoftwo
 \else \expandafter \@secondoftwo
 \fi
}%
\providecommand \natexlab [1]{#1}%
\providecommand \enquote  [1]{``#1''}%
\providecommand \bibnamefont  [1]{#1}%
\providecommand \bibfnamefont [1]{#1}%
\providecommand \citenamefont [1]{#1}%
\providecommand \href@noop [0]{\@secondoftwo}%
\providecommand \href [0]{\begingroup \@sanitize@url \@href}%
\providecommand \@href[1]{\@@startlink{#1}\@@href}%
\providecommand \@@href[1]{\endgroup#1\@@endlink}%
\providecommand \@sanitize@url [0]{\catcode `\\12\catcode `\$12\catcode
  `\&12\catcode `\#12\catcode `\^12\catcode `\_12\catcode `\%12\relax}%
\providecommand \@@startlink[1]{}%
\providecommand \@@endlink[0]{}%
\providecommand \url  [0]{\begingroup\@sanitize@url \@url }%
\providecommand \@url [1]{\endgroup\@href {#1}{\urlprefix }}%
\providecommand \urlprefix  [0]{URL }%
\providecommand \Eprint [0]{\href }%
\providecommand \doibase [0]{http://dx.doi.org/}%
\providecommand \selectlanguage [0]{\@gobble}%
\providecommand \bibinfo  [0]{\@secondoftwo}%
\providecommand \bibfield  [0]{\@secondoftwo}%
\providecommand \translation [1]{[#1]}%
\providecommand \BibitemOpen [0]{}%
\providecommand \bibitemStop [0]{}%
\providecommand \bibitemNoStop [0]{.\EOS\space}%
\providecommand \EOS [0]{\spacefactor3000\relax}%
\providecommand \BibitemShut  [1]{\csname bibitem#1\endcsname}%
\let\auto@bib@innerbib\@empty
\bibitem [{\citenamefont {Van~der Pol}\ and\ \citenamefont {Van
  Der~Mark}(1927)}]{van1927frequency}%
  \BibitemOpen
  \bibfield  {author} {\bibinfo {author} {\bibfnamefont {B.}~\bibnamefont
  {Van~der Pol}}\ and\ \bibinfo {author} {\bibfnamefont {J.}~\bibnamefont {Van
  Der~Mark}},\ }\href@noop {} {\bibfield  {journal} {\bibinfo  {journal}
  {Nature}\ }\textbf {\bibinfo {volume} {120}},\ \bibinfo {pages} {363}
  (\bibinfo {year} {1927})}\BibitemShut {NoStop}%
\bibitem [{\citenamefont {May}(1976)}]{may1976simple}%
  \BibitemOpen
  \bibfield  {author} {\bibinfo {author} {\bibfnamefont {R.~M.}\ \bibnamefont
  {May}},\ }\href@noop {} {\bibfield  {journal} {\bibinfo  {journal} {Nature}\
  }\textbf {\bibinfo {volume} {261}},\ \bibinfo {pages} {459} (\bibinfo {year}
  {1976})}\BibitemShut {NoStop}%
\bibitem [{\citenamefont {Cross}\ and\ \citenamefont
  {Hohenberg}(1993)}]{CrossHohenberg}%
  \BibitemOpen
  \bibfield  {author} {\bibinfo {author} {\bibfnamefont {M.~C.}\ \bibnamefont
  {Cross}}\ and\ \bibinfo {author} {\bibfnamefont {P.~C.}\ \bibnamefont
  {Hohenberg}},\ }\href {\doibase 10.1103/RevModPhys.65.851} {\bibfield
  {journal} {\bibinfo  {journal} {Rev. Mod. Phys.}\ }\textbf {\bibinfo {volume}
  {65}},\ \bibinfo {pages} {851} (\bibinfo {year} {1993})}\BibitemShut
  {NoStop}%
\bibitem [{\citenamefont {Brown}\ \emph {et~al.}(1984)\citenamefont {Brown},
  \citenamefont {Mozurkewich},\ and\ \citenamefont {Gr\"uner}}]{Brown1984}%
  \BibitemOpen
  \bibfield  {author} {\bibinfo {author} {\bibfnamefont {S.~E.}\ \bibnamefont
  {Brown}}, \bibinfo {author} {\bibfnamefont {G.}~\bibnamefont {Mozurkewich}},
  \ and\ \bibinfo {author} {\bibfnamefont {G.}~\bibnamefont {Gr\"uner}},\
  }\href {\doibase 10.1103/PhysRevLett.52.2277} {\bibfield  {journal} {\bibinfo
   {journal} {Phys. Rev. Lett.}\ }\textbf {\bibinfo {volume} {52}},\ \bibinfo
  {pages} {2277} (\bibinfo {year} {1984})}\BibitemShut {NoStop}%
\bibitem [{\citenamefont {Parlitz}\ \emph {et~al.}(1997)\citenamefont
  {Parlitz}, \citenamefont {Junge},\ and\ \citenamefont
  {Kocarev}}]{parlitz1997subharmonic}%
  \BibitemOpen
  \bibfield  {author} {\bibinfo {author} {\bibfnamefont {U.}~\bibnamefont
  {Parlitz}}, \bibinfo {author} {\bibfnamefont {L.}~\bibnamefont {Junge}}, \
  and\ \bibinfo {author} {\bibfnamefont {L.}~\bibnamefont {Kocarev}},\
  }\href@noop {} {\bibfield  {journal} {\bibinfo  {journal} {Physical review
  letters}\ }\textbf {\bibinfo {volume} {79}},\ \bibinfo {pages} {3158}
  (\bibinfo {year} {1997})}\BibitemShut {NoStop}%
\bibitem [{\citenamefont {Rasband}(2015)}]{rasband2015chaotic}%
  \BibitemOpen
  \bibfield  {author} {\bibinfo {author} {\bibfnamefont {S.~N.}\ \bibnamefont
  {Rasband}},\ }\href@noop {} {\emph {\bibinfo {title} {Chaotic dynamics of
  nonlinear systems}}}\ (\bibinfo  {publisher} {Courier Dover Publications},\
  \bibinfo {year} {2015})\BibitemShut {NoStop}%
\bibitem [{\citenamefont {Strogatz}(2014)}]{strogatz2014nonlinear}%
  \BibitemOpen
  \bibfield  {author} {\bibinfo {author} {\bibfnamefont {S.~H.}\ \bibnamefont
  {Strogatz}},\ }\href@noop {} {\emph {\bibinfo {title} {Nonlinear dynamics and
  chaos: with applications to physics, biology, chemistry, and engineering}}}\
  (\bibinfo  {publisher} {Westview press},\ \bibinfo {year} {2014})\BibitemShut
  {NoStop}%
\bibitem [{\citenamefont {Linsay}(1981)}]{linsay1981period}%
  \BibitemOpen
  \bibfield  {author} {\bibinfo {author} {\bibfnamefont {P.~S.}\ \bibnamefont
  {Linsay}},\ }\href@noop {} {\bibfield  {journal} {\bibinfo  {journal}
  {Physical Review Letters}\ }\textbf {\bibinfo {volume} {47}},\ \bibinfo
  {pages} {1349} (\bibinfo {year} {1981})}\BibitemShut {NoStop}%
\bibitem [{\citenamefont {Kaneko}(1984)}]{kaneko1984period}%
  \BibitemOpen
  \bibfield  {author} {\bibinfo {author} {\bibfnamefont {K.}~\bibnamefont
  {Kaneko}},\ }\href@noop {} {\bibfield  {journal} {\bibinfo  {journal}
  {Progress of Theoretical Physics}\ }\textbf {\bibinfo {volume} {72}},\
  \bibinfo {pages} {480} (\bibinfo {year} {1984})}\BibitemShut {NoStop}%
\bibitem [{\citenamefont {Jackson}\ and\ \citenamefont
  {H{\"u}bler}(1990)}]{jackson1990periodic}%
  \BibitemOpen
  \bibfield  {author} {\bibinfo {author} {\bibfnamefont {E.~A.}\ \bibnamefont
  {Jackson}}\ and\ \bibinfo {author} {\bibfnamefont {A.}~\bibnamefont
  {H{\"u}bler}},\ }\href@noop {} {\bibfield  {journal} {\bibinfo  {journal}
  {Physica D: Nonlinear Phenomena}\ }\textbf {\bibinfo {volume} {44}},\
  \bibinfo {pages} {407} (\bibinfo {year} {1990})}\BibitemShut {NoStop}%
\bibitem [{Note1()}]{Note1}%
  \BibitemOpen
  \bibinfo {note} {The reader may object that the distinction between the
  driven and un-driven case is spurious, since we may always add an additional
  coordinate $\partial _t \lambda = \omega _{\protect \textrm {D}}$ and
  consider the periodic drive to be a coupling to this coordinate. However,
  this entails an all-to-one interaction, in contrast to the requirement of
  local couplings, so we feel the distinction is worth preserving.}\BibitemShut
  {Stop}%
\bibitem [{\citenamefont {Van~der Pol}(1926)}]{van1926}%
  \BibitemOpen
  \bibfield  {author} {\bibinfo {author} {\bibfnamefont {B.}~\bibnamefont
  {Van~der Pol}},\ }\href@noop {} {\bibfield  {journal} {\bibinfo  {journal}
  {The London, Edinburgh, and Dublin Philosophical Magazine and Journal of
  Science}\ }\textbf {\bibinfo {volume} {2}},\ \bibinfo {pages} {978} (\bibinfo
  {year} {1926})}\BibitemShut {NoStop}%
\bibitem [{\citenamefont {Kuramoto}(1975)}]{kuramoto1975self}%
  \BibitemOpen
  \bibfield  {author} {\bibinfo {author} {\bibfnamefont {Y.}~\bibnamefont
  {Kuramoto}},\ }in\ \href@noop {} {\emph {\bibinfo {booktitle} {International
  symposium on mathematical problems in theoretical physics}}}\ (\bibinfo
  {organization} {Springer},\ \bibinfo {year} {1975})\ pp.\ \bibinfo {pages}
  {420--422}\BibitemShut {NoStop}%
\bibitem [{Note2()}]{Note2}%
  \BibitemOpen
  \bibinfo {note} {To define period doubling in certain systems (in particular,
  for closed Hamiltonian dynamics) some notion of coarse graining may be
  required. Indeed, any specific solution may be aperiodic, but an appropriate
  ensemble average over the initial state, spatial position, and/or time will
  reveal that the dynamics are period doubled in a \protect \emph {statistical}
  sense.}\BibitemShut {Stop}%
\bibitem [{\citenamefont {Brown}\ \emph {et~al.}(1985)\citenamefont {Brown},
  \citenamefont {Mozurkewich},\ and\ \citenamefont
  {Gr{\"u}ner}}]{brown1985harmonic}%
  \BibitemOpen
  \bibfield  {author} {\bibinfo {author} {\bibfnamefont {S.~E.}\ \bibnamefont
  {Brown}}, \bibinfo {author} {\bibfnamefont {G.}~\bibnamefont {Mozurkewich}},
  \ and\ \bibinfo {author} {\bibfnamefont {G.}~\bibnamefont {Gr{\"u}ner}},\
  }\href@noop {} {\bibfield  {journal} {\bibinfo  {journal} {Solid state
  communications}\ }\textbf {\bibinfo {volume} {54}},\ \bibinfo {pages} {23}
  (\bibinfo {year} {1985})}\BibitemShut {NoStop}%
\bibitem [{\citenamefont {Tua}\ and\ \citenamefont
  {Ruvalds}(1985)}]{tua1985dynamics}%
  \BibitemOpen
  \bibfield  {author} {\bibinfo {author} {\bibfnamefont {P.}~\bibnamefont
  {Tua}}\ and\ \bibinfo {author} {\bibfnamefont {J.}~\bibnamefont {Ruvalds}},\
  }\href@noop {} {\bibfield  {journal} {\bibinfo  {journal} {Solid state
  communications}\ }\textbf {\bibinfo {volume} {54}},\ \bibinfo {pages} {471}
  (\bibinfo {year} {1985})}\BibitemShut {NoStop}%
\bibitem [{\citenamefont {Sherwin}\ and\ \citenamefont
  {Zettl}(1985)}]{sherwin1985complete}%
  \BibitemOpen
  \bibfield  {author} {\bibinfo {author} {\bibfnamefont {M.}~\bibnamefont
  {Sherwin}}\ and\ \bibinfo {author} {\bibfnamefont {A.}~\bibnamefont
  {Zettl}},\ }\href@noop {} {\bibfield  {journal} {\bibinfo  {journal}
  {Physical Review B}\ }\textbf {\bibinfo {volume} {32}},\ \bibinfo {pages}
  {5536} (\bibinfo {year} {1985})}\BibitemShut {NoStop}%
\bibitem [{\citenamefont {Wiesenfeld}\ and\ \citenamefont
  {Satija}(1987)}]{Wiesenfeld1987}%
  \BibitemOpen
  \bibfield  {author} {\bibinfo {author} {\bibfnamefont {K.}~\bibnamefont
  {Wiesenfeld}}\ and\ \bibinfo {author} {\bibfnamefont {I.}~\bibnamefont
  {Satija}},\ }\href {\doibase 10.1103/PhysRevB.36.2483} {\bibfield  {journal}
  {\bibinfo  {journal} {Phys. Rev. B}\ }\textbf {\bibinfo {volume} {36}},\
  \bibinfo {pages} {2483} (\bibinfo {year} {1987})}\BibitemShut {NoStop}%
\bibitem [{\citenamefont {Bhattacharya}\ \emph {et~al.}(1987)\citenamefont
  {Bhattacharya}, \citenamefont {Stokes}, \citenamefont {Higgins},\ and\
  \citenamefont {Klemm}}]{Bhattacharya1987}%
  \BibitemOpen
  \bibfield  {author} {\bibinfo {author} {\bibfnamefont {S.}~\bibnamefont
  {Bhattacharya}}, \bibinfo {author} {\bibfnamefont {J.~P.}\ \bibnamefont
  {Stokes}}, \bibinfo {author} {\bibfnamefont {M.~J.}\ \bibnamefont {Higgins}},
  \ and\ \bibinfo {author} {\bibfnamefont {R.~A.}\ \bibnamefont {Klemm}},\
  }\href {\doibase 10.1103/PhysRevLett.59.1849} {\bibfield  {journal} {\bibinfo
   {journal} {Phys. Rev. Lett.}\ }\textbf {\bibinfo {volume} {59}},\ \bibinfo
  {pages} {1849} (\bibinfo {year} {1987})}\BibitemShut {NoStop}%
\bibitem [{\citenamefont {Falo}\ \emph {et~al.}(1993)\citenamefont {Falo},
  \citenamefont {Flor\'{\i}a}, \citenamefont {Mart\'{\i}nez},\ and\
  \citenamefont {Mazo}}]{Falo93}%
  \BibitemOpen
  \bibfield  {author} {\bibinfo {author} {\bibfnamefont {F.}~\bibnamefont
  {Falo}}, \bibinfo {author} {\bibfnamefont {L.~M.}\ \bibnamefont
  {Flor\'{\i}a}}, \bibinfo {author} {\bibfnamefont {P.~J.}\ \bibnamefont
  {Mart\'{\i}nez}}, \ and\ \bibinfo {author} {\bibfnamefont {J.~J.}\
  \bibnamefont {Mazo}},\ }\href {\doibase 10.1103/PhysRevB.48.7434} {\bibfield
  {journal} {\bibinfo  {journal} {Phys. Rev. B}\ }\textbf {\bibinfo {volume}
  {48}},\ \bibinfo {pages} {7434} (\bibinfo {year} {1993})}\BibitemShut
  {NoStop}%
\bibitem [{\citenamefont {Balents}\ and\ \citenamefont
  {Fisher}(1995)}]{balents1995temporal}%
  \BibitemOpen
  \bibfield  {author} {\bibinfo {author} {\bibfnamefont {L.}~\bibnamefont
  {Balents}}\ and\ \bibinfo {author} {\bibfnamefont {M.~P.}\ \bibnamefont
  {Fisher}},\ }\href@noop {} {\bibfield  {journal} {\bibinfo  {journal}
  {Physical review letters}\ }\textbf {\bibinfo {volume} {75}},\ \bibinfo
  {pages} {4270} (\bibinfo {year} {1995})}\BibitemShut {NoStop}%
\bibitem [{\citenamefont {Lee}\ \emph {et~al.}(1991)\citenamefont {Lee},
  \citenamefont {Newrock}, \citenamefont {Mast}, \citenamefont {Hebboul},
  \citenamefont {Garland},\ and\ \citenamefont {Lobb}}]{Lee1991}%
  \BibitemOpen
  \bibfield  {author} {\bibinfo {author} {\bibfnamefont {H.~C.}\ \bibnamefont
  {Lee}}, \bibinfo {author} {\bibfnamefont {R.~S.}\ \bibnamefont {Newrock}},
  \bibinfo {author} {\bibfnamefont {D.~B.}\ \bibnamefont {Mast}}, \bibinfo
  {author} {\bibfnamefont {S.~E.}\ \bibnamefont {Hebboul}}, \bibinfo {author}
  {\bibfnamefont {J.~C.}\ \bibnamefont {Garland}}, \ and\ \bibinfo {author}
  {\bibfnamefont {C.~J.}\ \bibnamefont {Lobb}},\ }\href {\doibase
  10.1103/PhysRevB.44.921} {\bibfield  {journal} {\bibinfo  {journal} {Phys.
  Rev. B}\ }\textbf {\bibinfo {volume} {44}},\ \bibinfo {pages} {921} (\bibinfo
  {year} {1991})}\BibitemShut {NoStop}%
\bibitem [{\citenamefont {Yu}\ \emph {et~al.}(1992)\citenamefont {Yu},
  \citenamefont {Harris}, \citenamefont {Hebboul}, \citenamefont {Garland},\
  and\ \citenamefont {Stroud}}]{Yu1992}%
  \BibitemOpen
  \bibfield  {author} {\bibinfo {author} {\bibfnamefont {W.}~\bibnamefont
  {Yu}}, \bibinfo {author} {\bibfnamefont {E.~B.}\ \bibnamefont {Harris}},
  \bibinfo {author} {\bibfnamefont {S.~E.}\ \bibnamefont {Hebboul}}, \bibinfo
  {author} {\bibfnamefont {J.~C.}\ \bibnamefont {Garland}}, \ and\ \bibinfo
  {author} {\bibfnamefont {D.}~\bibnamefont {Stroud}},\ }\href {\doibase
  10.1103/PhysRevB.45.12624} {\bibfield  {journal} {\bibinfo  {journal} {Phys.
  Rev. B}\ }\textbf {\bibinfo {volume} {45}},\ \bibinfo {pages} {12624}
  (\bibinfo {year} {1992})}\BibitemShut {NoStop}%
\bibitem [{\citenamefont {Wilczek}(2012)}]{wilczek2012quantum}%
  \BibitemOpen
  \bibfield  {author} {\bibinfo {author} {\bibfnamefont {F.}~\bibnamefont
  {Wilczek}},\ }\href@noop {} {\bibfield  {journal} {\bibinfo  {journal}
  {Physical review letters}\ }\textbf {\bibinfo {volume} {109}},\ \bibinfo
  {pages} {160401} (\bibinfo {year} {2012})}\BibitemShut {NoStop}%
\bibitem [{\citenamefont {Shapere}\ and\ \citenamefont
  {Wilczek}(2012)}]{shapere2012classical}%
  \BibitemOpen
  \bibfield  {author} {\bibinfo {author} {\bibfnamefont {A.}~\bibnamefont
  {Shapere}}\ and\ \bibinfo {author} {\bibfnamefont {F.}~\bibnamefont
  {Wilczek}},\ }\href@noop {} {\bibfield  {journal} {\bibinfo  {journal}
  {Physical review letters}\ }\textbf {\bibinfo {volume} {109}},\ \bibinfo
  {pages} {160402} (\bibinfo {year} {2012})}\BibitemShut {NoStop}%
\bibitem [{\citenamefont {Bruno}(2013)}]{bruno2013impossibility}%
  \BibitemOpen
  \bibfield  {author} {\bibinfo {author} {\bibfnamefont {P.}~\bibnamefont
  {Bruno}},\ }\href@noop {} {\bibfield  {journal} {\bibinfo  {journal}
  {Physical review letters}\ }\textbf {\bibinfo {volume} {111}},\ \bibinfo
  {pages} {070402} (\bibinfo {year} {2013})}\BibitemShut {NoStop}%
\bibitem [{\citenamefont {Nozi{\`e}res}(2013)}]{nozieres2013time}%
  \BibitemOpen
  \bibfield  {author} {\bibinfo {author} {\bibfnamefont {P.}~\bibnamefont
  {Nozi{\`e}res}},\ }\href@noop {} {\bibfield  {journal} {\bibinfo  {journal}
  {EPL (Europhysics Letters)}\ }\textbf {\bibinfo {volume} {103}},\ \bibinfo
  {pages} {57008} (\bibinfo {year} {2013})}\BibitemShut {NoStop}%
\bibitem [{\citenamefont {Volovik}(2013)}]{volovik2013broken}%
  \BibitemOpen
  \bibfield  {author} {\bibinfo {author} {\bibfnamefont {G.~E.}\ \bibnamefont
  {Volovik}},\ }\href@noop {} {\bibfield  {journal} {\bibinfo  {journal} {JETP
  letters}\ }\textbf {\bibinfo {volume} {98}},\ \bibinfo {pages} {491}
  (\bibinfo {year} {2013})}\BibitemShut {NoStop}%
\bibitem [{\citenamefont {Sacha}(2015)}]{sacha2015modeling}%
  \BibitemOpen
  \bibfield  {author} {\bibinfo {author} {\bibfnamefont {K.}~\bibnamefont
  {Sacha}},\ }\href@noop {} {\bibfield  {journal} {\bibinfo  {journal}
  {Physical Review A}\ }\textbf {\bibinfo {volume} {91}},\ \bibinfo {pages}
  {033617} (\bibinfo {year} {2015})}\BibitemShut {NoStop}%
\bibitem [{\citenamefont {Watanabe}\ and\ \citenamefont
  {Oshikawa}(2015)}]{watanabe2015absence}%
  \BibitemOpen
  \bibfield  {author} {\bibinfo {author} {\bibfnamefont {H.}~\bibnamefont
  {Watanabe}}\ and\ \bibinfo {author} {\bibfnamefont {M.}~\bibnamefont
  {Oshikawa}},\ }\href@noop {} {\bibfield  {journal} {\bibinfo  {journal}
  {Physical review letters}\ }\textbf {\bibinfo {volume} {114}},\ \bibinfo
  {pages} {251603} (\bibinfo {year} {2015})}\BibitemShut {NoStop}%
\bibitem [{Note3()}]{Note3}%
  \BibitemOpen
  \bibinfo {note} {In certain special cases \cite
  {citro2015dynamical,Chandran2016}, it may be possible for a driven, many-body
  system to avoid its ergodic fate \cite {liggett2012interacting}.}\BibitemShut
  {Stop}%
\bibitem [{\citenamefont {Khemani}\ \emph {et~al.}(2016)\citenamefont
  {Khemani}, \citenamefont {Lazarides}, \citenamefont {Moessner},\ and\
  \citenamefont {Sondhi}}]{khemani2016phase}%
  \BibitemOpen
  \bibfield  {author} {\bibinfo {author} {\bibfnamefont {V.}~\bibnamefont
  {Khemani}}, \bibinfo {author} {\bibfnamefont {A.}~\bibnamefont {Lazarides}},
  \bibinfo {author} {\bibfnamefont {R.}~\bibnamefont {Moessner}}, \ and\
  \bibinfo {author} {\bibfnamefont {S.~L.}\ \bibnamefont {Sondhi}},\
  }\href@noop {} {\bibfield  {journal} {\bibinfo  {journal} {Physical review
  letters}\ }\textbf {\bibinfo {volume} {116}},\ \bibinfo {pages} {250401}
  (\bibinfo {year} {2016})}\BibitemShut {NoStop}%
\bibitem [{\citenamefont {Else}\ \emph {et~al.}(2016)\citenamefont {Else},
  \citenamefont {Bauer},\ and\ \citenamefont {Nayak}}]{else2016floquet}%
  \BibitemOpen
  \bibfield  {author} {\bibinfo {author} {\bibfnamefont {D.~V.}\ \bibnamefont
  {Else}}, \bibinfo {author} {\bibfnamefont {B.}~\bibnamefont {Bauer}}, \ and\
  \bibinfo {author} {\bibfnamefont {C.}~\bibnamefont {Nayak}},\ }\href@noop {}
  {\bibfield  {journal} {\bibinfo  {journal} {Physical review letters}\
  }\textbf {\bibinfo {volume} {117}},\ \bibinfo {pages} {090402} (\bibinfo
  {year} {2016})}\BibitemShut {NoStop}%
\bibitem [{\citenamefont {von Keyserlingk}\ \emph {et~al.}(2016)\citenamefont
  {von Keyserlingk}, \citenamefont {Khemani},\ and\ \citenamefont
  {Sondhi}}]{von2016absolute}%
  \BibitemOpen
  \bibfield  {author} {\bibinfo {author} {\bibfnamefont {C.~W.}\ \bibnamefont
  {von Keyserlingk}}, \bibinfo {author} {\bibfnamefont {V.}~\bibnamefont
  {Khemani}}, \ and\ \bibinfo {author} {\bibfnamefont {S.~L.}\ \bibnamefont
  {Sondhi}},\ }\href@noop {} {\bibfield  {journal} {\bibinfo  {journal}
  {Physical Review B}\ }\textbf {\bibinfo {volume} {94}},\ \bibinfo {pages}
  {085112} (\bibinfo {year} {2016})}\BibitemShut {NoStop}%
\bibitem [{\citenamefont {Yao}\ \emph {et~al.}(2017)\citenamefont {Yao},
  \citenamefont {Potter}, \citenamefont {Potirniche},\ and\ \citenamefont
  {Vishwanath}}]{yao2017discrete}%
  \BibitemOpen
  \bibfield  {author} {\bibinfo {author} {\bibfnamefont {N.~Y.}\ \bibnamefont
  {Yao}}, \bibinfo {author} {\bibfnamefont {A.~C.}\ \bibnamefont {Potter}},
  \bibinfo {author} {\bibfnamefont {I.-D.}\ \bibnamefont {Potirniche}}, \ and\
  \bibinfo {author} {\bibfnamefont {A.}~\bibnamefont {Vishwanath}},\
  }\href@noop {} {\bibfield  {journal} {\bibinfo  {journal} {Physical Review
  Letters}\ }\textbf {\bibinfo {volume} {118}},\ \bibinfo {pages} {030401}
  (\bibinfo {year} {2017})}\BibitemShut {NoStop}%
\bibitem [{\citenamefont {Khemani}\ \emph {et~al.}(2017)\citenamefont
  {Khemani}, \citenamefont {von Keyserlingk},\ and\ \citenamefont
  {Sondhi}}]{Khemani2017}%
  \BibitemOpen
  \bibfield  {author} {\bibinfo {author} {\bibfnamefont {V.}~\bibnamefont
  {Khemani}}, \bibinfo {author} {\bibfnamefont {C.~W.}\ \bibnamefont {von
  Keyserlingk}}, \ and\ \bibinfo {author} {\bibfnamefont {S.~L.}\ \bibnamefont
  {Sondhi}},\ }\href {\doibase 10.1103/PhysRevB.96.115127} {\bibfield
  {journal} {\bibinfo  {journal} {Phys. Rev. B}\ }\textbf {\bibinfo {volume}
  {96}},\ \bibinfo {pages} {115127} (\bibinfo {year} {2017})}\BibitemShut
  {NoStop}%
\bibitem [{\citenamefont {Lazarides}\ and\ \citenamefont
  {Moessner}(2017)}]{LazaridesMoessner2017}%
  \BibitemOpen
  \bibfield  {author} {\bibinfo {author} {\bibfnamefont {A.}~\bibnamefont
  {Lazarides}}\ and\ \bibinfo {author} {\bibfnamefont {R.}~\bibnamefont
  {Moessner}},\ }\href {\doibase 10.1103/PhysRevB.95.195135} {\bibfield
  {journal} {\bibinfo  {journal} {Phys. Rev. B}\ }\textbf {\bibinfo {volume}
  {95}},\ \bibinfo {pages} {195135} (\bibinfo {year} {2017})}\BibitemShut
  {NoStop}%
\bibitem [{\citenamefont {Zhang}\ \emph {et~al.}(2017)\citenamefont {Zhang},
  \citenamefont {Hess}, \citenamefont {Kyprianidis}, \citenamefont {Becker},
  \citenamefont {Lee}, \citenamefont {Smith}, \citenamefont {Pagano},
  \citenamefont {Potirniche}, \citenamefont {Potter}, \citenamefont
  {Vishwanath} \emph {et~al.}}]{zhang2017observation}%
  \BibitemOpen
  \bibfield  {author} {\bibinfo {author} {\bibfnamefont {J.}~\bibnamefont
  {Zhang}}, \bibinfo {author} {\bibfnamefont {P.}~\bibnamefont {Hess}},
  \bibinfo {author} {\bibfnamefont {A.}~\bibnamefont {Kyprianidis}}, \bibinfo
  {author} {\bibfnamefont {P.}~\bibnamefont {Becker}}, \bibinfo {author}
  {\bibfnamefont {A.}~\bibnamefont {Lee}}, \bibinfo {author} {\bibfnamefont
  {J.}~\bibnamefont {Smith}}, \bibinfo {author} {\bibfnamefont
  {G.}~\bibnamefont {Pagano}}, \bibinfo {author} {\bibfnamefont {I.-D.}\
  \bibnamefont {Potirniche}}, \bibinfo {author} {\bibfnamefont {A.~C.}\
  \bibnamefont {Potter}}, \bibinfo {author} {\bibfnamefont {A.}~\bibnamefont
  {Vishwanath}},  \emph {et~al.},\ }\href@noop {} {\bibfield  {journal}
  {\bibinfo  {journal} {Nature}\ }\textbf {\bibinfo {volume} {543}},\ \bibinfo
  {pages} {217} (\bibinfo {year} {2017})}\BibitemShut {NoStop}%
\bibitem [{\citenamefont {Choi}\ \emph {et~al.}(2017)\citenamefont {Choi},
  \citenamefont {Choi}, \citenamefont {Landig}, \citenamefont {Kucsko},
  \citenamefont {Zhou}, \citenamefont {Isoya}, \citenamefont {Jelezko},
  \citenamefont {Onoda}, \citenamefont {Sumiya}, \citenamefont {Khemani} \emph
  {et~al.}}]{choi2017observation}%
  \BibitemOpen
  \bibfield  {author} {\bibinfo {author} {\bibfnamefont {S.}~\bibnamefont
  {Choi}}, \bibinfo {author} {\bibfnamefont {J.}~\bibnamefont {Choi}}, \bibinfo
  {author} {\bibfnamefont {R.}~\bibnamefont {Landig}}, \bibinfo {author}
  {\bibfnamefont {G.}~\bibnamefont {Kucsko}}, \bibinfo {author} {\bibfnamefont
  {H.}~\bibnamefont {Zhou}}, \bibinfo {author} {\bibfnamefont {J.}~\bibnamefont
  {Isoya}}, \bibinfo {author} {\bibfnamefont {F.}~\bibnamefont {Jelezko}},
  \bibinfo {author} {\bibfnamefont {S.}~\bibnamefont {Onoda}}, \bibinfo
  {author} {\bibfnamefont {H.}~\bibnamefont {Sumiya}}, \bibinfo {author}
  {\bibfnamefont {V.}~\bibnamefont {Khemani}},  \emph {et~al.},\ }\href@noop {}
  {\bibfield  {journal} {\bibinfo  {journal} {Nature}\ }\textbf {\bibinfo
  {volume} {543}},\ \bibinfo {pages} {221} (\bibinfo {year}
  {2017})}\BibitemShut {NoStop}%
\bibitem [{\citenamefont {Nyquist}(1928)}]{nyquist1928thermal}%
  \BibitemOpen
  \bibfield  {author} {\bibinfo {author} {\bibfnamefont {H.}~\bibnamefont
  {Nyquist}},\ }\href@noop {} {\bibfield  {journal} {\bibinfo  {journal}
  {Physical review}\ }\textbf {\bibinfo {volume} {32}},\ \bibinfo {pages} {110}
  (\bibinfo {year} {1928})}\BibitemShut {NoStop}%
\bibitem [{\citenamefont {Wolfram}(1983)}]{wolfram1983statistical}%
  \BibitemOpen
  \bibfield  {author} {\bibinfo {author} {\bibfnamefont {S.}~\bibnamefont
  {Wolfram}},\ }\href@noop {} {\bibfield  {journal} {\bibinfo  {journal}
  {Reviews of modern physics}\ }\textbf {\bibinfo {volume} {55}},\ \bibinfo
  {pages} {601} (\bibinfo {year} {1983})}\BibitemShut {NoStop}%
\bibitem [{\citenamefont {G{\'a}cs}(1986)}]{gacs1986reliable}%
  \BibitemOpen
  \bibfield  {author} {\bibinfo {author} {\bibfnamefont {P.}~\bibnamefont
  {G{\'a}cs}},\ }\href@noop {} {\bibfield  {journal} {\bibinfo  {journal}
  {Journal of Computer and System Sciences}\ }\textbf {\bibinfo {volume}
  {32}},\ \bibinfo {pages} {15} (\bibinfo {year} {1986})}\BibitemShut {NoStop}%
\bibitem [{\citenamefont {G{\'a}cs}(2001)}]{gacs2001reliable}%
  \BibitemOpen
  \bibfield  {author} {\bibinfo {author} {\bibfnamefont {P.}~\bibnamefont
  {G{\'a}cs}},\ }\href@noop {} {\bibfield  {journal} {\bibinfo  {journal}
  {Journal of Statistical Physics}\ }\textbf {\bibinfo {volume} {103}},\
  \bibinfo {pages} {45} (\bibinfo {year} {2001})}\BibitemShut {NoStop}%
\bibitem [{\citenamefont {Gray}(2001)}]{gray2001reader}%
  \BibitemOpen
  \bibfield  {author} {\bibinfo {author} {\bibfnamefont {L.~F.}\ \bibnamefont
  {Gray}},\ }\href@noop {} {\bibfield  {journal} {\bibinfo  {journal} {Journal
  of Statistical Physics}\ }\textbf {\bibinfo {volume} {103}},\ \bibinfo
  {pages} {1} (\bibinfo {year} {2001})}\BibitemShut {NoStop}%
\bibitem [{\citenamefont {Toom}(1974)}]{toom1974nonergodic}%
  \BibitemOpen
  \bibfield  {author} {\bibinfo {author} {\bibfnamefont {A.~L.}\ \bibnamefont
  {Toom}},\ }\href@noop {} {\bibfield  {journal} {\bibinfo  {journal} {Problemy
  Peredachi Informatsii}\ }\textbf {\bibinfo {volume} {10}},\ \bibinfo {pages}
  {70} (\bibinfo {year} {1974})}\BibitemShut {NoStop}%
\bibitem [{\citenamefont {Toom}(1976)}]{toom1976unstable}%
  \BibitemOpen
  \bibfield  {author} {\bibinfo {author} {\bibfnamefont {A.}~\bibnamefont
  {Toom}},\ }\href@noop {} {\bibfield  {journal} {\bibinfo  {journal} {Problemy
  Peredachi Informatsii}\ }\textbf {\bibinfo {volume} {12}},\ \bibinfo {pages}
  {78} (\bibinfo {year} {1976})}\BibitemShut {NoStop}%
\bibitem [{\citenamefont {Toom}(1980)}]{toom1980multicomponent}%
  \BibitemOpen
  \bibfield  {author} {\bibinfo {author} {\bibfnamefont {A.}~\bibnamefont
  {Toom}},\ }\href@noop {} {\  (\bibinfo {year} {1980})}\BibitemShut {NoStop}%
\bibitem [{\citenamefont {Bennett}\ \emph {et~al.}(1990)\citenamefont
  {Bennett}, \citenamefont {Grinstein}, \citenamefont {He}, \citenamefont
  {Jayaprakash},\ and\ \citenamefont {Mukamel}}]{bennett1990stability}%
  \BibitemOpen
  \bibfield  {author} {\bibinfo {author} {\bibfnamefont {C.~H.}\ \bibnamefont
  {Bennett}}, \bibinfo {author} {\bibfnamefont {G.}~\bibnamefont {Grinstein}},
  \bibinfo {author} {\bibfnamefont {Y.}~\bibnamefont {He}}, \bibinfo {author}
  {\bibfnamefont {C.}~\bibnamefont {Jayaprakash}}, \ and\ \bibinfo {author}
  {\bibfnamefont {D.}~\bibnamefont {Mukamel}},\ }\href@noop {} {\bibfield
  {journal} {\bibinfo  {journal} {Physical Review A}\ }\textbf {\bibinfo
  {volume} {41}},\ \bibinfo {pages} {1932} (\bibinfo {year}
  {1990})}\BibitemShut {NoStop}%
\bibitem [{\citenamefont {Landau}\ and\ \citenamefont
  {Lifshitz}(1969)}]{LandauStatmech}%
  \BibitemOpen
  \bibfield  {author} {\bibinfo {author} {\bibfnamefont {L.~D.}\ \bibnamefont
  {Landau}}\ and\ \bibinfo {author} {\bibfnamefont {E.~M.}\ \bibnamefont
  {Lifshitz}},\ }\href@noop {} {\emph {\bibinfo {title} {Statistical Physics
  V5: Course of Theoretical Physics}}}\ (\bibinfo  {publisher} {Pergamon
  press},\ \bibinfo {year} {1969})\BibitemShut {NoStop}%
\bibitem [{\citenamefont {Frenkel}\ and\ \citenamefont
  {Kontorova}(1938)}]{FK38}%
  \BibitemOpen
  \bibfield  {author} {\bibinfo {author} {\bibfnamefont {J.}~\bibnamefont
  {Frenkel}}\ and\ \bibinfo {author} {\bibfnamefont {T.}~\bibnamefont
  {Kontorova}},\ }\href@noop {} {\bibfield  {journal} {\bibinfo  {journal}
  {Physik. Z. Sowietunion}\ ,\ \bibinfo {pages} {1}} (\bibinfo {year}
  {1938})}\BibitemShut {NoStop}%
\bibitem [{\citenamefont {Braun}\ and\ \citenamefont
  {Kivshar}(2013)}]{braun2013frenkel}%
  \BibitemOpen
  \bibfield  {author} {\bibinfo {author} {\bibfnamefont {O.~M.}\ \bibnamefont
  {Braun}}\ and\ \bibinfo {author} {\bibfnamefont {Y.~S.}\ \bibnamefont
  {Kivshar}},\ }\href@noop {} {\emph {\bibinfo {title} {The Frenkel-Kontorova
  model: concepts, methods, and applications}}}\ (\bibinfo  {publisher}
  {Springer Science \& Business Media},\ \bibinfo {year} {2013})\BibitemShut
  {NoStop}%
\bibitem [{Note4()}]{Note4}%
  \BibitemOpen
  \bibinfo {note} {Such a model can be realized in many different experimental
  systems. One simple circuit QED architecture is as follows. Consider a linear
  array of superconducting islands, with each island coupled to a ground line
  via a Josephson junction with a shunt capacitance. To modulate the Josephson
  coupling, one can utilize a flux-tuned split junction with an oscillatory
  flux generated from a nearby AC current loop. Finally, the interaction term
  can be obtained by coupling each island to its neighbor with a
  ``superinductor''.}\BibitemShut {Stop}%
\bibitem [{sup()}]{suppinfo}%
  \BibitemOpen
  \href@noop {} {\bibinfo  {journal} {See supplementary information}\
  }\BibitemShut {NoStop}%
\bibitem [{\citenamefont {Zounes}\ and\ \citenamefont
  {Rand}(2002)}]{Zounes2002}%
  \BibitemOpen
\bibfield  {journal} {  }\bibfield  {author} {\bibinfo {author} {\bibfnamefont
  {R.~S.}\ \bibnamefont {Zounes}}\ and\ \bibinfo {author} {\bibfnamefont
  {R.~H.}\ \bibnamefont {Rand}},\ }\href {\doibase
  http://dx.doi.org/10.1016/S0020-7462(00)00095-0} {\bibfield  {journal}
  {\bibinfo  {journal} {International Journal of Non-Linear Mechanics}\
  }\textbf {\bibinfo {volume} {37}},\ \bibinfo {pages} {43 } (\bibinfo {year}
  {2002})}\BibitemShut {NoStop}%
\bibitem [{\citenamefont {Brizard}(2013)}]{Brizard2013}%
  \BibitemOpen
  \bibfield  {author} {\bibinfo {author} {\bibfnamefont {A.~J.}\ \bibnamefont
  {Brizard}},\ }\href {\doibase http://dx.doi.org/10.1016/j.cnsns.2012.08.023}
  {\bibfield  {journal} {\bibinfo  {journal} {Communications in Nonlinear
  Science and Numerical Simulation}\ }\textbf {\bibinfo {volume} {18}},\
  \bibinfo {pages} {511 } (\bibinfo {year} {2013})}\BibitemShut {NoStop}%
\bibitem [{Note5()}]{Note5}%
  \BibitemOpen
  \bibinfo {note} {In fact, the $q \to -q$ symmetry of the original Hamiltonian
  is inessential to the period-doubling physics we discuss, analogous to
  quantum case. \cite {von2016absolute}}\BibitemShut {NoStop}%
\bibitem [{\citenamefont {Else}\ \emph {et~al.}(2017)\citenamefont {Else},
  \citenamefont {Bauer},\ and\ \citenamefont {Nayak}}]{ElseBauerNayak}%
  \BibitemOpen
  \bibfield  {author} {\bibinfo {author} {\bibfnamefont {D.~V.}\ \bibnamefont
  {Else}}, \bibinfo {author} {\bibfnamefont {B.}~\bibnamefont {Bauer}}, \ and\
  \bibinfo {author} {\bibfnamefont {C.}~\bibnamefont {Nayak}},\ }\href
  {\doibase 10.1103/PhysRevX.7.011026} {\bibfield  {journal} {\bibinfo
  {journal} {Phys. Rev. X}\ }\textbf {\bibinfo {volume} {7}},\ \bibinfo {pages}
  {011026} (\bibinfo {year} {2017})}\BibitemShut {NoStop}%
\bibitem [{\citenamefont {Stewart}(1968)}]{stewart1968current}%
  \BibitemOpen
  \bibfield  {author} {\bibinfo {author} {\bibfnamefont {W.}~\bibnamefont
  {Stewart}},\ }\href@noop {} {\bibfield  {journal} {\bibinfo  {journal}
  {Applied Physics Letters}\ }\textbf {\bibinfo {volume} {12}},\ \bibinfo
  {pages} {277} (\bibinfo {year} {1968})}\BibitemShut {NoStop}%
\bibitem [{\citenamefont {McCumber}(1968)}]{mccumber1968effect}%
  \BibitemOpen
  \bibfield  {author} {\bibinfo {author} {\bibfnamefont {D.}~\bibnamefont
  {McCumber}},\ }\href@noop {} {\bibfield  {journal} {\bibinfo  {journal}
  {Journal of Applied Physics}\ }\textbf {\bibinfo {volume} {39}},\ \bibinfo
  {pages} {3113} (\bibinfo {year} {1968})}\BibitemShut {NoStop}%
\bibitem [{\citenamefont {B\"uttiker}\ \emph {et~al.}(1983)\citenamefont
  {B\"uttiker}, \citenamefont {Harris},\ and\ \citenamefont
  {Landauer}}]{buttiker83}%
  \BibitemOpen
  \bibfield  {author} {\bibinfo {author} {\bibfnamefont {M.}~\bibnamefont
  {B\"uttiker}}, \bibinfo {author} {\bibfnamefont {E.~P.}\ \bibnamefont
  {Harris}}, \ and\ \bibinfo {author} {\bibfnamefont {R.}~\bibnamefont
  {Landauer}},\ }\href {\doibase 10.1103/PhysRevB.28.1268} {\bibfield
  {journal} {\bibinfo  {journal} {Phys. Rev. B}\ }\textbf {\bibinfo {volume}
  {28}},\ \bibinfo {pages} {1268} (\bibinfo {year} {1983})}\BibitemShut
  {NoStop}%
\bibitem [{Note6()}]{Note6}%
  \BibitemOpen
  \bibinfo {note} {Roughly speaking, if we perturb about the period-doubled
  solution of a single oscillator $q(t)$, $q_j(t) = q(t) + \Delta q_j(t)$, in
  the limit $\delta \to 0$ the dispersion relation of the perturbation is
  strongly off-resonant with the drive, $\omega _{\protect \textrm {eff}}(k) /
  \omega _{\protect \textrm {D}}\sim \delta $, so the finite damping $\eta $
  prevents heating and the system settles into a period-doubled steady
  state.}\BibitemShut {Stop}%
\bibitem [{\citenamefont {Braun}\ \emph
  {et~al.}(1997{\natexlab{a}})\citenamefont {Braun}, \citenamefont {Dauxois},
  \citenamefont {Paliy},\ and\ \citenamefont {Peyrard}}]{braun1997nonlinear}%
  \BibitemOpen
  \bibfield  {author} {\bibinfo {author} {\bibfnamefont {O.~M.}\ \bibnamefont
  {Braun}}, \bibinfo {author} {\bibfnamefont {T.}~\bibnamefont {Dauxois}},
  \bibinfo {author} {\bibfnamefont {M.~V.}\ \bibnamefont {Paliy}}, \ and\
  \bibinfo {author} {\bibfnamefont {M.}~\bibnamefont {Peyrard}},\ }\href@noop
  {} {\bibfield  {journal} {\bibinfo  {journal} {Physical Review E}\ }\textbf
  {\bibinfo {volume} {55}},\ \bibinfo {pages} {3598} (\bibinfo {year}
  {1997}{\natexlab{a}})}\BibitemShut {NoStop}%
\bibitem [{\citenamefont {Braun}\ \emph
  {et~al.}(1997{\natexlab{b}})\citenamefont {Braun}, \citenamefont {Bishop},\
  and\ \citenamefont {R{\"o}der}}]{braun1997hysteresis}%
  \BibitemOpen
  \bibfield  {author} {\bibinfo {author} {\bibfnamefont {O.}~\bibnamefont
  {Braun}}, \bibinfo {author} {\bibfnamefont {A.}~\bibnamefont {Bishop}}, \
  and\ \bibinfo {author} {\bibfnamefont {J.}~\bibnamefont {R{\"o}der}},\
  }\href@noop {} {\bibfield  {journal} {\bibinfo  {journal} {Physical review
  letters}\ }\textbf {\bibinfo {volume} {79}},\ \bibinfo {pages} {3692}
  (\bibinfo {year} {1997}{\natexlab{b}})}\BibitemShut {NoStop}%
\bibitem [{Note7()}]{Note7}%
  \BibitemOpen
  \bibinfo {note} {To test this hypothesis, we can change the effective mass to
  be positive by changing the non-linearity $\epsilon $ of the pendulum (e.g.,
  by replacing $\protect \qopname \relax o{cos}(q) \to \protect \frac {1}{2}
  q^2 + \protect \frac {1}{24} q^4$, giving $\epsilon = -\protect \frac
  {1}{8}$). The period doubled solutions now exist at minima of $\protect
  \mathaccentV {bar}016{H}$, corresponding to positive temperature, and
  analogous simulations indeed show that the pendula now synchronize
  ferromagnetically.}\BibitemShut {Stop}%
\bibitem [{\citenamefont {Kapitza}(1951)}]{kapitza1951}%
  \BibitemOpen
  \bibfield  {author} {\bibinfo {author} {\bibfnamefont {P.~L.}\ \bibnamefont
  {Kapitza}},\ }\href@noop {} {\bibfield  {journal} {\bibinfo  {journal}
  {Uspekhi Fizicheskikh Nauk}\ }\textbf {\bibinfo {volume} {44}},\ \bibinfo
  {pages} {7} (\bibinfo {year} {1951})}\BibitemShut {NoStop}%
\bibitem [{dhu()}]{dhuse}%
  \BibitemOpen
  \href@noop {} {\ }\bibinfo {note} {We thank D. Huse for discussions on this
  point.}\BibitemShut {Stop}%
\bibitem [{\citenamefont {Huse}\ and\ \citenamefont
  {Fisher}(1987)}]{huse1987dynamics}%
  \BibitemOpen
  \bibfield  {author} {\bibinfo {author} {\bibfnamefont {D.~A.}\ \bibnamefont
  {Huse}}\ and\ \bibinfo {author} {\bibfnamefont {D.~S.}\ \bibnamefont
  {Fisher}},\ }\href@noop {} {\bibfield  {journal} {\bibinfo  {journal}
  {Physical Review B}\ }\textbf {\bibinfo {volume} {35}},\ \bibinfo {pages}
  {6841} (\bibinfo {year} {1987})}\BibitemShut {NoStop}%
\bibitem [{Note8()}]{Note8}%
  \BibitemOpen
  \bibinfo {note} {An analogous analysis would apply to a continuous-time
  Fourier transform.}\BibitemShut {Stop}%
\bibitem [{\citenamefont {Welch}(1967)}]{welch1967use}%
  \BibitemOpen
  \bibfield  {author} {\bibinfo {author} {\bibfnamefont {P.}~\bibnamefont
  {Welch}},\ }\href@noop {} {\bibfield  {journal} {\bibinfo  {journal} {IEEE
  Transactions on audio and electroacoustics}\ }\textbf {\bibinfo {volume}
  {15}},\ \bibinfo {pages} {70} (\bibinfo {year} {1967})}\BibitemShut {NoStop}%
\bibitem [{\citenamefont {Liggett}(2012)}]{liggett2012interacting}%
  \BibitemOpen
  \bibfield  {author} {\bibinfo {author} {\bibfnamefont {T.~M.}\ \bibnamefont
  {Liggett}},\ }\href@noop {} {\emph {\bibinfo {title} {Interacting particle
  systems}}},\ Vol.\ \bibinfo {volume} {276}\ (\bibinfo  {publisher}
  {Springer-Verlag},\ \bibinfo {year} {2012})\BibitemShut {NoStop}%
\bibitem [{\citenamefont {Grinstein}\ \emph {et~al.}(1993)\citenamefont
  {Grinstein}, \citenamefont {Mukamel}, \citenamefont {Seidin},\ and\
  \citenamefont {Bennett}}]{grinstein1993temporally}%
  \BibitemOpen
  \bibfield  {author} {\bibinfo {author} {\bibfnamefont {G.}~\bibnamefont
  {Grinstein}}, \bibinfo {author} {\bibfnamefont {D.}~\bibnamefont {Mukamel}},
  \bibinfo {author} {\bibfnamefont {R.}~\bibnamefont {Seidin}}, \ and\ \bibinfo
  {author} {\bibfnamefont {C.~H.}\ \bibnamefont {Bennett}},\ }\href@noop {}
  {\bibfield  {journal} {\bibinfo  {journal} {Physical review letters}\
  }\textbf {\bibinfo {volume} {70}},\ \bibinfo {pages} {3607} (\bibinfo {year}
  {1993})}\BibitemShut {NoStop}%
\bibitem [{\citenamefont {Citro}\ \emph {et~al.}(2015)\citenamefont {Citro},
  \citenamefont {Dalla~Torre}, \citenamefont {D?Alessio}, \citenamefont
  {Polkovnikov}, \citenamefont {Babadi}, \citenamefont {Oka},\ and\
  \citenamefont {Demler}}]{citro2015dynamical}%
  \BibitemOpen
  \bibfield  {author} {\bibinfo {author} {\bibfnamefont {R.}~\bibnamefont
  {Citro}}, \bibinfo {author} {\bibfnamefont {E.~G.}\ \bibnamefont
  {Dalla~Torre}}, \bibinfo {author} {\bibfnamefont {L.}~\bibnamefont
  {D?Alessio}}, \bibinfo {author} {\bibfnamefont {A.}~\bibnamefont
  {Polkovnikov}}, \bibinfo {author} {\bibfnamefont {M.}~\bibnamefont {Babadi}},
  \bibinfo {author} {\bibfnamefont {T.}~\bibnamefont {Oka}}, \ and\ \bibinfo
  {author} {\bibfnamefont {E.}~\bibnamefont {Demler}},\ }\href@noop {}
  {\bibfield  {journal} {\bibinfo  {journal} {Annals of Physics}\ }\textbf
  {\bibinfo {volume} {360}},\ \bibinfo {pages} {694} (\bibinfo {year}
  {2015})}\BibitemShut {NoStop}%
\bibitem [{\citenamefont {Abanin}\ \emph {et~al.}(2017)\citenamefont {Abanin},
  \citenamefont {De~Roeck}, \citenamefont {Ho},\ and\ \citenamefont
  {Huveneers}}]{Abanin2017}%
  \BibitemOpen
  \bibfield  {author} {\bibinfo {author} {\bibfnamefont {D.~A.}\ \bibnamefont
  {Abanin}}, \bibinfo {author} {\bibfnamefont {W.}~\bibnamefont {De~Roeck}},
  \bibinfo {author} {\bibfnamefont {W.~W.}\ \bibnamefont {Ho}}, \ and\ \bibinfo
  {author} {\bibfnamefont {F.~m.~c.}\ \bibnamefont {Huveneers}},\ }\href
  {\doibase 10.1103/PhysRevB.95.014112} {\bibfield  {journal} {\bibinfo
  {journal} {Phys. Rev. B}\ }\textbf {\bibinfo {volume} {95}},\ \bibinfo
  {pages} {014112} (\bibinfo {year} {2017})}\BibitemShut {NoStop}%
\bibitem [{\citenamefont {Martin}\ \emph {et~al.}(1973)\citenamefont {Martin},
  \citenamefont {Siggia},\ and\ \citenamefont {Rose}}]{MartinSiggiaRose}%
  \BibitemOpen
  \bibfield  {author} {\bibinfo {author} {\bibfnamefont {P.~C.}\ \bibnamefont
  {Martin}}, \bibinfo {author} {\bibfnamefont {E.~D.}\ \bibnamefont {Siggia}},
  \ and\ \bibinfo {author} {\bibfnamefont {H.~A.}\ \bibnamefont {Rose}},\
  }\href {\doibase 10.1103/PhysRevA.8.423} {\bibfield  {journal} {\bibinfo
  {journal} {Phys. Rev. A}\ }\textbf {\bibinfo {volume} {8}},\ \bibinfo {pages}
  {423} (\bibinfo {year} {1973})}\BibitemShut {NoStop}%
\bibitem [{\citenamefont {Chandran}\ and\ \citenamefont
  {Sondhi}(2016)}]{Chandran2016}%
  \BibitemOpen
  \bibfield  {author} {\bibinfo {author} {\bibfnamefont {A.}~\bibnamefont
  {Chandran}}\ and\ \bibinfo {author} {\bibfnamefont {S.~L.}\ \bibnamefont
  {Sondhi}},\ }\href {\doibase 10.1103/PhysRevB.93.174305} {\bibfield
  {journal} {\bibinfo  {journal} {Phys. Rev. B}\ }\textbf {\bibinfo {volume}
  {93}},\ \bibinfo {pages} {174305} (\bibinfo {year} {2016})}\BibitemShut
  {NoStop}%
\end{thebibliography}%

\onecolumngrid
\appendix

\end{document}


\title{Supplemental Material for Classical Discrete Time Crystals}
\author{N. Y. Yao, C. Nayak, L. Balents, M. P. Zaletel}
\maketitle

\section{Langevin integrator}
\label{app:langevin}
\noindent Numerical simulations were performed using a symplectic Langevin integrator. The equations of motion ($m=1$) read
\begin{align}
\dot{q} &= \frac{\partial H}{\partial p} \\
\dot{p} &= -  \frac{\partial H}{\partial q} - \eta p + \xi(t)
\end{align}
We consider the case $H = \frac{p^2}{2} + V(q, t)$, where we can Trotterize the evolution by interleaving the updates	
\begin{align}
q(t + \Delta t / 2) &= q(t) + p(t) \Delta t / 2 \\
p(t + \Delta t) &= F / \eta + e^{- \eta \Delta t} ( p(t) - F(t) / \eta) \\
q(t + \Delta t / 2) &= q(t) + p(t) \Delta t / 2
\end{align}
where $F(t) = -  \frac{\partial V}{\partial p}(q(t), t) + \xi$.  At each step, $\xi$ is a random variable evenly distributed in $[-\sqrt{6 T \eta}, \sqrt{6 T \eta}]$, which reproduces the variance $\langle \xi(t) \xi(t') \rangle = 2 \eta T \delta(t - t')$.
Throughout, the calculations were performed using $\Delta t =  \frac{1}{80} \frac{2 \pi}{\od}$.

\section{Effective DC-Hamiltonian for finite $u$.}
\label{app:finiteu}
\noindent In the rotating frame, the time-averaged single oscillator Hamiltonian is
\begin{align}
\bar{H} &= \delta (1 + u) J/4 - \frac{1}{2} \epsilon J^2 + \frac{\delta}{4} J \left( \cos(2 \tth) - 1 \right)
\label{eq:magnusSupp}
\end{align}
where $u = 4 (1 - \od / 2) / \delta$.

\vspace{2mm}

\noindent Dissipation and finite temperature enter through the Langevin dynamics $F_{\textrm{b}}(t) = - \eta \dot{q} + \xi(t)$, where $\xi(t)$ is a stochastic force  $\langle \xi(t) \xi(t') \rangle = 2 \eta T \delta(t - t')$. In the rotating frame, the contribution to the equations of motion are
	\begin{align}
\dot{\tilde{\theta}} & \owns q ( -\eta p + \xi) = - \eta J \sin(2 \tilde{\theta} +  \od t  ) + \xi \sqrt{2 J} \cos(\tilde{\theta} + \od t /2 ) \\
\dot{J}  & \owns  p ( -\eta p + \xi) = - 2 J \eta \sin^2(\tilde{\theta} + \od t / 2) + \xi \sqrt{2 J} \sin(\tilde{\theta} + \od  t /2 )
\end{align}	
If we average the dissipative force $-\eta \dot{q}$ over one period, we see it only effects the equations of motion for $J$. Thus, we approximate the effective of the Langevin force as
\begin{align}
\dot{\tilde{\theta}} &  \owns q ( -\eta p + \xi) \sim 0  \\
\dot{J}  & \owns  p ( -\eta p + \xi) \sim  -  J \eta + \sqrt{J} \xi
\end{align}	
Presumably it shouldn't matter that we've approximated the Langevin dynamics in this fashion, since we could have chosen to use a Langevin force for $J$ at the outset.

\vspace{2mm}

\noindent The equations of motion in the rotating frame are then:
\begin{align}
\dot{\tilde{\theta}} &= \delta (1 + u) /4 -  \epsilon J + \frac{\delta}{4} ( \cos( 2 \tilde{\theta}) - 1) \\
\dot{J}  &=  \frac{\delta}{2} \sin ( 2 \tilde{\theta}) J - \eta J + \sqrt{J} \xi(t).
\end{align}
At $T = 0$, the fixed point condition $\dot{J} = \dot{\tilde{\theta}} = 0$ is
\begin{align}
 \sin(2 \tilde{\theta}) & = F, \quad F = \frac{2 \eta}{\delta} \\
 \bar{J} &=  \frac{\delta}{4 \epsilon}( u  \pm \sqrt{1 - ( \frac{2 \eta}{\delta} )^2 }).
\end{align}
Implicitly this assumes $\bar{J} > 0$ (otherwise the solution is unphysical), so the condition for a period doubled resonance is
\begin{align}
|F| & < 1 \\
u &> -  \sqrt{1 - F^2 }.
\end{align}
These conditions define the phase boundaries of the main text. To arrive at a more familiar effective Hamiltonian in which no cross-terms between $\tilde{\theta}$ and $J$ appear, first define $\bar{J}(\tilde{\theta})$ by the condition $\dot{\tilde{\theta}} = 0$,
\begin{align}
\bar{J}(\tilde{\theta}) &=  \frac{\delta}{4 \epsilon}( u  + \cos( 2 \tilde{\theta}) )
\end{align}
We then define $\tilde{J} = J  - \bar{J}(\tilde{\theta})$, which is a canonical transformation. In these variables,
\begin{align}
\bar{H}[\tilde{\theta}, \tilde{J}] = - \frac{\epsilon}{2} \tilde{J}^2  + \frac{\delta^2}{16 \epsilon}( u \cos(2 \tilde{\theta}) + \cos^2(2 \tilde{\theta}) ) 
\end{align}
The effect of the Langevin bath is to add $\dot{\tilde{J}} \owns - \eta (\tilde{J} + \bar{J}(\tilde{\theta})) + \xi(t)$ to the equations of motion. This can be expressed as
\begin{align}
H_{\textrm{eff}} &=  \bar{H} + \eta \int \bar{J}(\tilde{\theta}) d \tilde{\theta} \\
\dot{\tilde{\theta}} &= \partial_{\tilde{J}} H_{\textrm{eff}}  \\
\dot{\tilde{J}} &= -\partial_{\tilde{\theta}} H_{\textrm{eff}}  - \eta  \tilde{J}  + \xi(t)
\end{align}
Apparently the effective Hamiltonian acquires a term
\begin{align}
\eta \int \bar{J}(\tilde{\theta}) d \tilde{\theta}  & = \frac{\delta}{4 \epsilon} \eta \left[ \tilde{\theta} u + \frac{1}{2} \sin (2 \tilde{\theta}) \right] \\
&= \eta \left[ \tilde{\theta} \frac{1 - \od/2}{\epsilon} + \frac{\delta}{8 \epsilon} \sin (2 \tilde{\theta}) \right] 
\end{align}
Due to the slope $ \frac{\delta}{4 \epsilon} \eta u \tilde{\theta}$, it is now a DC-driven model, and the problem is intrinsically non-equilibrium. 
In the fully sliding regime, the force $\frac{\delta}{4 \epsilon} \eta u$ is balanced by the friction $\eta / \epsilon$. In this limit we can equate the two, $\partial_t \tilde{\theta} = \frac{\delta}{4} u = 1 - \od/2$, or $\partial_t \theta = 1$ in the original frame. 

\vspace{2mm}

\noindent We now consider the coupling $H_g = - g \sum_i (q_i - q_{i+1})^2$. Time-averaging the coupling in the rotating frame,
\begin{align}
\bar{H}_g &= \frac{1}{\td} \frac{g}{2}  \sum_i \int_0^{\td} dt  \left (  \sqrt{2 J_i} \cos(\tth_i + \od t / 2) - \sqrt{2 J_{i+1}} \cos(\tth_{i+1} + \od t / 2)  \right)^2 \\
&= \frac{g}{2} \sum_i \left[ J_i + J_{i+1} - 2 \sqrt{J_i J_{i+1} } \cos(\tth_i - \tth_{i+1} ) \right]
\end{align}
Letting $J_i = \bar{J} + \tilde{J}_i$, and keeping terms to second order in $\tilde{J}_i, \tilde{\theta}_i - \tilde{\theta}_{i+1}$:
\begin{align}
\bar{H}_g \approx \frac{g}{2} \sum_i \frac{1}{4 \bar{J}}(\tilde{J}_i - \tilde{J}_{i+1})^2 - 2 \bar{J} \cos(\tth_i - \tth_{i+1} ) 
\end{align}
While there is no difficulty in keeping both terms,  in the large-$u$ limit the second term is larger by a factor of $u$, which is why we neglect the first in the main text.

\vspace{2mm}

\noindent In conclusion, we have mapped the parametrically AC-driven FK model to a \textbf{DC}-driven FK model.

\section{Validating the handedness of phase slips in the driven pendulum problem}
\label{app:handedness}
\noindent When the damping is $\bar{J} \eta$ is comparable to the washboard height $2 \bar{J}  \delta$, the effective model predicts the phase $\tilde{\theta}$ will begin rolling along the ramp - though because the mass is negative for $\epsilon > 0$,  it will actually go \emph{up} the ramp. To show this is not an artifact of the approximations used to derive the effective Hamiltonian (such as the Magnus expansion an averaging of the Langevin force), we numerically demonstrate this effect for the original parametrically driven oscillator coupled to a Langevin bath. 

\vspace{2mm}

\begin{figure}[t]
\includegraphics[width=0.6\columnwidth]{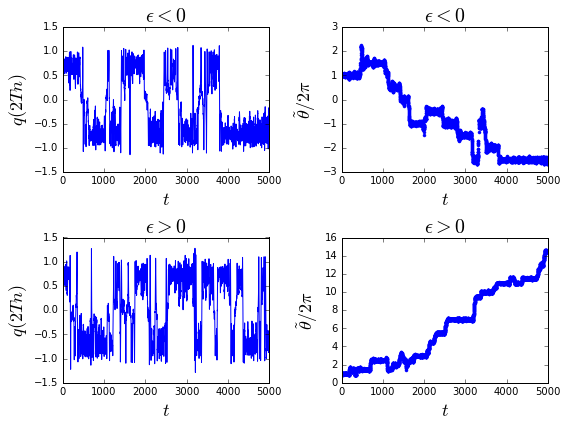}
\caption{Finite temperature drift in $\tilde{\theta}$ depending on the sign of the mass $\epsilon$ (which can be adjusted by using a different an anharmonicity, $F(q) = q \pm q^3 / 6 + \cdots$. Note that when looking at $q(2 \td n)$, it is impossible to distinguish the two cases - they just look like $\pi$-phase slips. But $\tilde{\theta}$ reveals the clear difference in the handedness.}
\label{fig:climb}
\end{figure}

\noindent  In Fig.\ref{fig:climb}, we show finite-$T$ dynamics of a single oscillator for a single noise realization,  both for $\epsilon = \frac{1}{8}$ (a $\cos(q)$ pendula) or  $\epsilon = -\frac{1}{8}$ (a $\frac{1}{2}q^2 + \frac{1}{4} q^4$ oscillator). The dynamics indeed show a biased random walk of opposite average velocity. In fact, for large $\bar{J} \eta$ this is not surprising. Assuming the linear ramp dominates over the washboard oscillations, the equations of motion for any $\epsilon$ read $\dot{\tilde{\theta}} = - \epsilon \tilde{J}, \dot{\tilde{J}} = - \eta \tilde{J} - (1- \od/2 ) / \epsilon$. At steady state, $\dot{\tilde{\theta}} = 1 - \od/2$, which by definition corresponds to the natural frequency $\dot{\theta} = 1$.

\bibliography{TCBib}